\documentclass[final]{svjour2}
\usepackage{graphicx}
\usepackage{rotating}
\usepackage{amssymb}
\usepackage{mathptmx}
\usepackage[numbers]{natbib}
\usepackage{multirow}
\usepackage{longtable}
\usepackage{bm}
\usepackage{amsmath}
\usepackage{amssymb}
\usepackage{epstopdf}
\usepackage{color}
\usepackage{ulem}
\makeatletter
\journalname{Journal of low temperature physics}
\bibpunct{}{}{,}{s}{}{,}

\begin{document}

\newcommand{\hdblarrow}{H\makebox[0.9ex][l]{$\downdownarrows$}-}

% Annotations made by SJY
\newcommand{\sjy}[1]{\textcolor{red}{#1}}

\title{Global Phase Diagram of the Kondo Lattice: From Heavy Fermion
Metals to Kondo Insulators}

\author{Seiji J. Yamamoto \and Qimiao Si}

\institute{
Seiji J. Yamamoto \at NHMFL and Department of Physics, Florida State University, \\
Tallahassee, Florida 32306, USA \\
\email{sjyamamoto@magnet.fsu.edu
\and
Qimiao Si \at Department of Physics and Astronomy, Rice University,\\ 
Houston, TX 77005, USA\\ 
Tel.: +1-713-348-5204\\ 
Fax: +1-713-348-4150\\
\email{qmsi@rice.edu}
}
}

%\date{24.06.2010}
\date{J. Low Temp. Phys. {\bf 161}, 233-262 (2010)}

\maketitle


\begin{abstract}
We discuss the general theoretical arguments advanced earlier for
the $T=0$ global phase diagram of antiferromagnetic Kondo lattice systems,
distinguishing between the established and the conjectured. 
In addition to the well-known phase of a paramagnetic metal with
a ``large'' Fermi surface (${\rm P_L}$), 
there is also an antiferromagnetic phase with a ``small'' Fermi surface 
(${\rm AF_S}$).
We provide the details of the derivation of 
a quantum non-linear sigma-model (QNL$\sigma$M) representation 
of the Kondo lattice Hamiltonian, which leads to an 
effective field theory containing both low-energy fermions in the 
vicinity of a Fermi surface and low-energy bosons near zero momentum.
An asymptotically exact analysis of this effective field theory
is made possible through the development of a renormalization
group procedure for mixed fermion-boson systems.
Considerations on how to connect the ${\rm AF_S}$ and ${\rm P_L}$
phases lead to a global phase diagram, which not only
puts into perspective the theory of local quantum criticality
for antiferromagnetic heavy fermion metals, but also provides the basis
to understand the surprising recent experiments
in chemically-doped as well as pressurized ${\rm YbRh_2Si_2}$.
We point out that the ${\rm AF_S}$ phase still occurs for the case
of an equal number of spin-$1/2$ local moments and conduction
electrons. This 
observation
raises the prospect for a global
phase diagram of heavy fermion systems in the Kondo-insulator regime.
Finally, we discuss the connection between the Kondo breakdown physics
discussed here for the Kondo lattice systems and the non-Fermi liquid
behavior recently studied from a holographic perspective.

PACS numbers: 
\end{abstract}

\section{Introduction}
Critical phenomenon associated with a second-order phase transition is 
formulated in terms of Landau's notion of spontaneous symmetry breaking.
Consider an antiferromagnetic transition of a rotationally-invariant
spin system. Above the N\'eel 
transition temperature, $T>T_N$, spins
retain the rotational invariance. Below the N\'eel temperature, $T<T_N$,
the rotational invariance is spontaneously broken, leading to an 
antiferromagnetic (AF) state in which the order parameter -- 
the staggered magnetization, ${\bf m}$ -- acquires a finite value.
Approaching the critical point, the order parameter vanishes but
strong spatial fluctuations of the order parameter remain.
% near the critical point. 
Indeed, 
%${\bf m} ({\bf x}) $ 
%characterizes the criticality,
the criticality is characterized by this fluctuating order parameter,
${\bf m} ({\bf x}) $,
% leading 
and is described by a 
Ginzburg-Landau theory in $d$-spatial dimensions.

What happens when the N\'eel temperature is tuned to zero? Hertz formulated
the problem in the context of itinerant systems, in which the order 
parameter ${\bf m}$ is associated with a microscopic magnetic state 
known as the
spin density wave (SDW) \cite{Hertz1976}. Hertz further assumed that 
the fluctuations 
of this SDW order parameter 
are
the only 
critical mode. For any quantum 
statistical system with Hamiltonian $H$, its partition function, 
$Z = {\rm Tr e}^{-\beta H}$, where $\beta \equiv 1/kT$, can be considered 
as 
a
quantum mechanical evolution operator along 
imaginary 
time $-i \tau$, 
where
$\tau$ 
is
defined in the range $(0,\hbar \beta)$ 
with
an associated periodic boundary condition. When $T_N$ is
nonzero,
finite-size effects
eventually cut
off the fluctuations along the imaginary time axis leaving classical critical
behavior associated with the spatial fluctuations of the order parameter.
When
$T_N \rightarrow 0$, on the other hand, the imaginary time axis has an
infinite extent. Correspondingly, the critical theory is a quantum 
Ginzburg-Landau theory of ${\bf m} ({\bf x}, \tau)$ in $d+z$ dimensions,
where $z$ is the dynamic exponent. The temporal fluctuations keep track
of the quantum nature of the collective fluctuations, but the degree
of freedom itself remains to be ${\bf m}$, the coarse-grained staggered 
magnetization, which is a classical variable. 

In recent 
years, 
it has become possible to test this picture 
of order-parameter fluctuations with
the explicit observation 
of the magnetic quantum critical points (QCPs) 
in heavy
fermion metals \cite{Stewart00,HvL-RMP,Gegenwart08}.
At the same time, a large body of theoretical work has
emerged on the quantum phase transitions
in the Kondo lattice Hamiltonian.

The purpose of this paper is multi-fold. We will first discuss 
the early theoretical and experimental motivations for 
the local quantum criticality with 
a 
Kondo breakdown. 
We describe the aspects in which this 
type of QCP differs from the Landau picture of order-parameter 
fluctuations, as well as the extensive theoretical developments 
and the pertinent experimental results.
The emphasis here is how the well-established heavy fermion state,
a paramagnetic metal with a ``large'' Fermi surface (${\rm P_L}$), 
can be critically broken down.
  
We proceed to consider the antiferromagnetically ordered part of the 
Kondo-lattice phase diagram. Our recent work \cite{Yamamoto07,Yamamoto08}
introduced a quantum non-linear sigma-model (QNL$\sigma$M) representation 
of the Kondo lattice Hamiltonian. An asymptotically exact
renormalization-group (RG) analysis establishes 
the existence of 
an antiferromagnetic phase with a ``small'' Fermi surface (${\rm AF_S}$).
Here, we provide the details of the derivation of this QNL$\sigma$M
representation, and  describe the RG procedure for this mixed 
fermion-boson system.

These considerations have lead to a proposal for a global phase 
diagram \cite{Si06,Yamamoto07, Si10}. This phase diagram puts the 
studies of 
local quantum criticality into a larger perspective.
In addition, recent experimental work on the Ir- and Co- doped 
${\rm YbRh_2Si_2}$ \cite{Friedemann09} have uncovered some surprising varieties
of quantum phase transitions in heavy fermion metals. Related observations
have appeared in pure ${\rm YbRh_2Si_2}$ under pressure \cite{Tokiwa.09} and 
Ge-doped ${\rm YbRh_2Si_2}$ \cite{Custers10}. These recent 
theoretical and experimental works have opened up a new direction for
the study of quantum criticality and novel phases in heavy fermion metals.
 
We also consider the case of 
heavy fermion systems in the regime 
of Kondo insulators. We show that 
the small-Fermi-surface 
antiferromagnetic phase, ${\rm AF_S}$, also exists in the 1+1 filling case.
Correspondingly, we propose that tuning a Kondo insulator towards larger
RKKY interactions can induce it into an antiferromagnetic or paramagnetic
metal with a small Fermi surface.

The focus of this article will be on the theoretical issues. 
More discussions on the experimental systems, with more references,
can be found in a recent article \cite{Si10} as well as in some 
of the comprehensive review articles \cite{Stewart00,HvL-RMP,Gegenwart08}.

\section{Local quantum criticality and Kondo breakdown from the paramagnetic
side}

\subsection{Kondo breakdown and local quantum criticality}
Early indications for the failure of the Hertz picture came from
inelastic neutron scattering measurements of the dynamical spin
susceptibility $\chi ({\bf q}, \omega)$. Following initial studies 
of dynamical scaling in heavy fermion metals \cite{Aronson95},
measurements of $\chi ({\bf q}, \omega)$ in the 
heavy fermion compound ${\rm CeCu_{6-x}Au_x}$
at the critical concentration $x_c \approx 0.1$, show that the
order-parameter fluctuations contain an anomalous critical exponent,
and display the property of $\omega/T$ scaling
\cite{Schroder2000,Schroder98,Stockert98}.
Both are properties of an interacting
fixed point. By contrast, the SDW QCP description 
\cite{Hertz1976,Millis93,Moriya1985}
of such systems
would have an effective dimensionality
%correspond to 
of $d+z \ge 4$, the upper critical dimension
of the quantum Ginzburg-Landau theory, and would correspond to 
a Gaussian fixed point with essentially mean-field exponents and 
violation of $\omega/T$ scaling.

These results motivated the theoretical proposal of a new class 
of QCPs. The key characteristic is the emergence 
of 
critical modes which are inherently quantum mechanical.
For the Kondo lattice system at hand, these modes are associated with
a critical Kondo breakdown. This Kondo breakdown picture was already
being considered prior to these experimental developments in some 
related theoretical models \cite{SiSmith96,SmithSi99,Sengupta00}.
In light of the observed interacting behavior, speculations were 
put forward \cite{SiSmithIngersent99,Coleman99}
that the Kondo breakdown underlies the interacting behavior
observed in the dynamical spin susceptibility. 
Soon thereafter, concrete theoretical formulations were 
advanced \cite{Si-Nature,Coleman01,Si-prb03}. 
The Kondo breakdown picture has subsequently been discussed in related
formulations \cite{Senthil04,PaulPepinNorman07}.

To put things in perspective, the traditional view of heavy-fermion metal\sjy{s}
is that the ground state is a Kondo singlet. This is a 
singlet state formed among all the local moments and all the conduction
electrons. This Kondo singlet state does not involve any spontaneous
symmetry breaking, but involves a macroscopic order. This is a quantum
entangled state. A continuous breakdown of the Kondo entanglement gives 
rise to the quantum critical modes.

A series of works
\cite{Si-Nature,Si-prb03,Grempel03,SunKotliar03,ZhuGrempelSi03,SunKotliar05, SiZhuGrempel05,Glossop07,Zhu07,Glossop_etal09} have studied the Kondo breakdown effect within an 
extended dynamical mean-field theory (EDMFT) 
\cite{SiSmith96,SmithSi00,Chitra00}
of the Kondo lattice Hamiltonian. The EDMFT equations have been studied 
using $\epsilon$-expansion RG procedure and various
numerical methods. When the Kondo breakdown occurs at the magnetic QCP,
the criticality depends on not only the order-parameter fluctuations 
but also the new Kondo-breakdown critical modes. Correspondingly, 
fractional critical exponents and $\omega/T$ scaling arise, in a way
that is consistent with the experimentally observed dynamical spin
susceptibility.

This local quantum critical picture also has implications for electronic 
excitations. Kondo resonances carry both spin ($1/2$) and charge ($e$). A Kondo
breakdown at the magnetic QCP leads to a jump of the 
Fermi surface at the magnetic quantum critical 
point \cite{SiSmithIngersent99,Coleman99,Si-Nature,Coleman01,Si-prb03},
as well as a Kondo-breakdown 
energy scale,
$E_{\rm loc}^*$, 
that continuously goes to zero
at the magnetic QCP. Both properties also operate 
when the Kondo breakdown occurs away from the magnetic 
QCP \cite{SiSmithIngersent99,Si-Nature,Si-prb03,Senthil04,PaulPepinNorman07}.

The Fermi surface jump has been probed extensively in several heavy-fermion metals.
In ${\rm YbRh_2Si_2}$, it has been demonstrated that Hall coefficient is dominated by the
normal component and therefore probes the Fermi surface \cite{paschen2004}. 
At low temperatures, the Hall coefficient displays a rapid crossover as a function of 
the non-thermal control parameter (a relatively small magnetic field). This crossover
extrapolates to a jump at the magnetic QCP
in the limit of zero temperature, providing evidence for a jump of the Fermi surface.
In ${\rm CeRhIn_5}$ \cite{park-nature06,Knebel08},
evidence for a Fermi surface jump accompanied by a mass divergence
has come from measurements of the de Haas-van Alphen effect \cite{shishido2005}.

The evidence for a Kondo-breakdown energy scale going to zero at the magnetic quantum
critical point has come from the measurements of both the Hall effect \cite{paschen2004}
and thermodynamic quantities \cite{Gegenwart2007}.

\subsection{Extended dynamical mean field theory of Kondo lattice}

The collapse of Kondo effect has been extensively studied from the
paramagnetic side.
%We wish to study 
A number of effects come into play: 
the RKKY interactions promote magnetic order,
the Kondo interactions favor Kondo-singlet formation,
%The relevant issues include the effect of 
%RKKY interactions,
%which induce magnetic order, and Kondo interactions, which 
%promote Kondo-singlet formation, as well as 
and the dynamical competition between these two types of interactions
are important for the transition.
One suitable microscopic approach
is the extended dynamical 
mean-field 
theory (EDMFT)~\cite{SiSmith96,SmithSi00,Chitra00}.
Here we outline the basic equations and the two type of solutions 
~\cite{Si-Nature,Si-prb03,Grempel03,ZhuGrempelSi03,SunKotliar03,Glossop07,Zhu07,Glossop_etal09}.
A more detailed summary can be found in Ref.~\cite{Si10}.

We consider the Kondo lattice model:
\begin{eqnarray}
	\mathcal{H} &=& \mathcal{H}_f +
\mathcal{H}_c +\mathcal{H}_K 
\label{KLM}
\end{eqnarray}
Here, $\mathcal{H}_c = \sum_{\vec{k}\sigma}\epsilon_{\vec{k}}
\psi^{\dagger}_{\vec{k}\sigma}
\psi_{\vec{k}\sigma}$ describes a band of free conduction $c-$electrons, 
with a bandwidth $W$.
$\mathcal{H}_K = \sum_{i} J_K {\vec S}_i \cdot {\vec s}_{c,i}$ 
specifies the
Kondo interaction of strength $J_K$;
here the conduction electron spin $\vec{s}_{c,i} = \frac{1}{2}
\sum_{\sigma\sigma^{\prime}}\psi^{\dagger}_{\sigma,i} 
\vec{\tau}_{\sigma\sigma^{\prime}}\psi_{\sigma^{\prime},i}$, 
where $\vec{\tau}$ is the vector of Pauli matrices.
Finally, $\mathcal{H}_f = \frac{1}{2}
\sum_{ij} I_{ij} {\vec S}_i \cdot {\vec S}_j$
is the magnetic Hamiltonian for the spin-$\frac{1}{2}$
$f-$moments, ${\vec S}_i$.
%,for which there is $1$ per site. 
The strength of the exchange interactions,
$I_{ij}$, is characterized by, say, the nearest neighbor value, $I$.

Within 
EDMFT, the dynamical spin susceptibility
and the conduction-electron Green's function
respectively have the forms
$\chi ({\bf q}, \omega) = [ I_{{\bf q}} + M(\omega) ]^{-1} $,
and
$G ({\bf k}, \varepsilon) =
[\varepsilon + \mu - \varepsilon_{\bf k} - \Sigma (\varepsilon)]^{-1} $.
The correlation functions,
$\chi ({\bf q}, \omega)$ and
$G ({\bf k}, \varepsilon)$, are momentum-dependent.
At the same time, the irreducible quantities,
$ M(\omega)$ and
$\Sigma (\varepsilon)$,
are momentum-independent.
They are determined in terms of
a
Bose-Fermi Kondo model,
\begin{eqnarray}
{\cal H}_{\text{imp}} &=& J_K ~{\bf S} \cdot {\bf s}_c +
\sum_{p,\sigma} E_{p}~c_{p\sigma}^{\dagger}~ c_{p\sigma}
%2col
%%\nonumber\\ &&
%2col
\nonumber \\
%%&& + \; g \sum_{p} {\bf S} ~\left( \pmb{\Phi}_{p} +
&& + \; g \sum_{p} {\bf S} \cdot \left( \pmb{\Phi}_{p} +
\pmb{\Phi}_{-p}^{\;\dagger} \right) +
%%\sum_{p}
\sum_{p}
w_{p}\,\pmb{\Phi}_{p}^{\;\dagger} \cdot \pmb{\Phi}_{p}\;.
% \nonumber\\
\label{QS:H-imp}
\end{eqnarray}
The fermionic ($c_{p\sigma}$) and bosonic 
($\pmb{\Phi}_{p}$)
baths are determined by self-consistency
conditions, which manifest the translational invariance,
%\begin{eqnarray}
%\chi_{{loc}}^a (\omega) &=& \sum_{\bf q} \chi^{a} ( {\bf q},
%\omega ) ,
%\nonumber \\[-1ex]
%\label{QS:self-consistent} \\[-1ex]
%G_{{loc}} (\omega) &=& \sum_{\bf k} G( {\bf k}, \omega )\;.
%\nonumber
%\end{eqnarray}
$\chi_{{loc}} (\omega)
= \sum_{\bf q} \chi ( {\bf q},
\omega )$,
and $G_{{loc}} (\varepsilon) = \sum_{\bf k} G( {\bf k}, \varepsilon )$.
%\;.
%\nonumber
%\end{eqnarray}
%When combined with the Dyson equations,
The 
$(0+1)$-dimensional 
quantum impurity problem,
Eq.~(\ref{QS:H-imp}), has the following Dyson equations:
$M(\omega)=\chi_{0}^{-1}(\omega) + 1/\chi_{\rm loc}(\omega)$
and $\Sigma(\varepsilon)=G_0^{-1}(\varepsilon) - 1/G_{\rm loc}(\varepsilon)$, where
$\chi_{0}^{-1} (\omega) = -g^2 \sum_p 2 w_{p} /(\omega^2 -
w_{p}^2)$
and $G_0 (\varepsilon) = \sum_p 1/(\varepsilon - E_p)$ are the
Weiss fields.
%, these self-consistency equations specify the
%dispersions, $E_{p}$ and $w_{p,a}$, and the coupling constant $g$.
The EDMFT formulation allows us to study different degrees of quantum
fluctuations as manifested in the spatial dimensionality of these
fluctuations. The case of two-dimensional
magnetic fluctuations are represented in terms of
the RKKY density of states that has a non-zero value at the lower edge,
e.g.,
\begin{eqnarray}
\rho_{I} (x) \equiv  \sum_{\bf q} \delta ( x  - I_{\bf q} )
= (1/{2 I}) \Theta(I - | x | ) \;,
\label{QS:rho_I_2D}
\end{eqnarray}
where $\Theta$ is the Heaviside step function.
%$\rho_{I} (\epsilon) \equiv  \sum_{\bf q} \delta ( \epsilon  - I_{\bf q} )
%= (1/{2 I}) \Theta(I - | \epsilon | )$.
Likewise, three-dimensional magnetic fluctuations are described
in terms of a $\rho_{I} (x)$ which vanishes at the lower edge
in a square-root fashion, 
e.g.,
\begin{eqnarray}
\rho_{I} (x) = (2/{\pi I^2}) \sqrt{I^2-x^2}\,
\Theta(I - | x | ) \;.
\label{QS:rho_I_3D}
\end{eqnarray}

The reduction of the Kondo-singlet amplitude
by the dynamical effects of 
RKKY interactions among the local
moments has been considered in some detail in a number of
studies based on EDMFT
\cite{Si-Nature,Si-prb03,Grempel03,ZhuGrempelSi03,SunKotliar03,Glossop07,Zhu07,Glossop_etal09}.
Irrespective of the spatial dimensionality, this weakening of the
Kondo effect is seen through the reduction of the $E_{\mathrm{loc}}^*$ scale.

Two classes of solutions emerge depending on whether this
Kondo breakdown scale vanishes at the AF QCP.
In the case of Eq.~(\ref{QS:rho_I_3D}), $E_{\mathrm{loc}}^*$ has not yet been
completely suppressed to zero when the AF QCP,
$\delta_c$, is reached from the paramagnetic
side.\footnote{However, it can still go to zero inside the AF
region \cite{SiSmithIngersent99,Si-Nature,Si-prb03}.}
The quantum critical behavior, at energies below $E_{\mathrm{loc}}^*$,
falls within the
SDW
type
\cite{Hertz1976,Millis93,Moriya1985}.
The zero-temperature dynamical spin susceptibility has the
following form:
\begin{eqnarray}
\chi({\bf q}, \omega ) =
\frac{1}{f({\bf q}) - ia \omega}
\;.
\label{QS:chi-qw-sdw}
\end{eqnarray}
Here $f({\bf q})=I_{\bf q}-I_{\bf Q}$, and is generically $\propto
({\bf q}-{\bf Q})^2 $ as the wavevector ${\bf q}$ approaches
the AF ordering wavevector ${\bf Q}$.
The QCP is described by a Gaussian fixed point. At non-zero temperatures,
a dangerously irrelevant operator invalidates the $\omega/T$
scaling~\cite{Millis93,Moriya1985}.

Another class of 
solutions 
corresponds to $E_{\mathrm{loc}}^*=0$ already at $\delta_c$.
It arises in the case of Eq.~(\ref{QS:rho_I_2D}), where the quantum critical
magnetic fluctuations are strong enough to suppress the Kondo effect.
The solution to the local spin susceptibility has the form
\begin{eqnarray}
\chi({\bf q}, \omega ) =
\frac{1}{f({\bf q}) + A \,(-i\omega)^{\alpha} W(\omega/T)}\;.
\label{QS:chi-qw-T}
\end{eqnarray}
This expression was derived~\cite{Si-Nature,Si-prb03}
within 
EDMFT studies,
through the aid of an $\epsilon$-expansion approach
to the Bose-Fermi Kondo model.
At the AF QCP,
the Kondo effect itself is critically destroyed.
The calculation of the critical exponent $\alpha$ is
beyond the reach of the $\epsilon$-expansion. In the
Ising-anisotropic case,
numerical calculations have found 
$\alpha\approx 0.7$~\cite{Grempel03,Glossop07,Zhu07,Glossop_etal09}.

The breakdown of the Kondo effect not only affects magnetic
dynamics, but also influences the single-electron excitations.
As the QCP is approached from the paramagnetic side, the 
quasi-particle
residue $z_L \propto (b^*)^2$,
where $b^*$ is the strength of a pole
in the conduction-electron self-energy $\Sigma$,
goes to zero. The large Fermi surface turns critical.

\section{Quantum non-linear sigma model representation of the Kondo lattice}

The local quantum criticality discussed in the previous section implies a breakdown of the 
Kondo effect at the magnetic QCP. In order to explore this issue further,
we study the Kondo effect deep inside the antiferromagnetic region of the phase diagram.
This regime arises when the Kondo coupling is infinitesimal compared to the AF exchange 
interaction among the local moments, {\it i.e.} in the regime
\begin{eqnarray}
J_K \ll I \ll W
\label{Kondo-AF-regime}
\end{eqnarray}
where $W$ is the width of the conduction electron band.

In the regime specified by Eq.~(\ref{Kondo-AF-regime}),
our strategy is to expand the Kondo lattice 
with respect to the limit of $J_K=0$, where the local moments and conduction electrons
decouple. The appropriate approach to implement this strategy comprises two steps.
First, 
we introduce a representation of the Kondo lattice Hamiltonian
in terms of a Quantum Nonlinear Sigma Model (QNL$\sigma$M). 
This representation, which results in  an effective low-energy field theory containing mixed fermions and bosons,
has been introduced in Ref.~\cite{Yamamoto07};
the goal of this section is to provide the details of the derivation.
Second, 
this effective low-energy theory is analyzed asymptotically exactly 
in terms of an RG method;
we will go through this in some detail in the following sections.
Related studies have also been pursued in Ref.~\cite{Ong_Jones09}.

%%%%%%%%%%%%
\subsection{Summary of the mapping}
%%%%%%%%%%%%

Since an explicit demonstration of the mapping will
take a fair amount of space, in this subsection we
summarize the essential points.
Subsequent subsections will provide the details.

The Kondo lattice Hamiltonian has been specified in Eq.~(\ref{KLM}).
For now, we will consider the electron concentration, $x$ per site,
to be
such that the Fermi surface
of $\mathcal{H}_c$ alone
does not touch the antiferromagnetic zone boundary.
Later, we will discuss the modifications necessary 
for the more general case.

\begin{figure}[hbtp]
   \centering
   \includegraphics[width=4in]{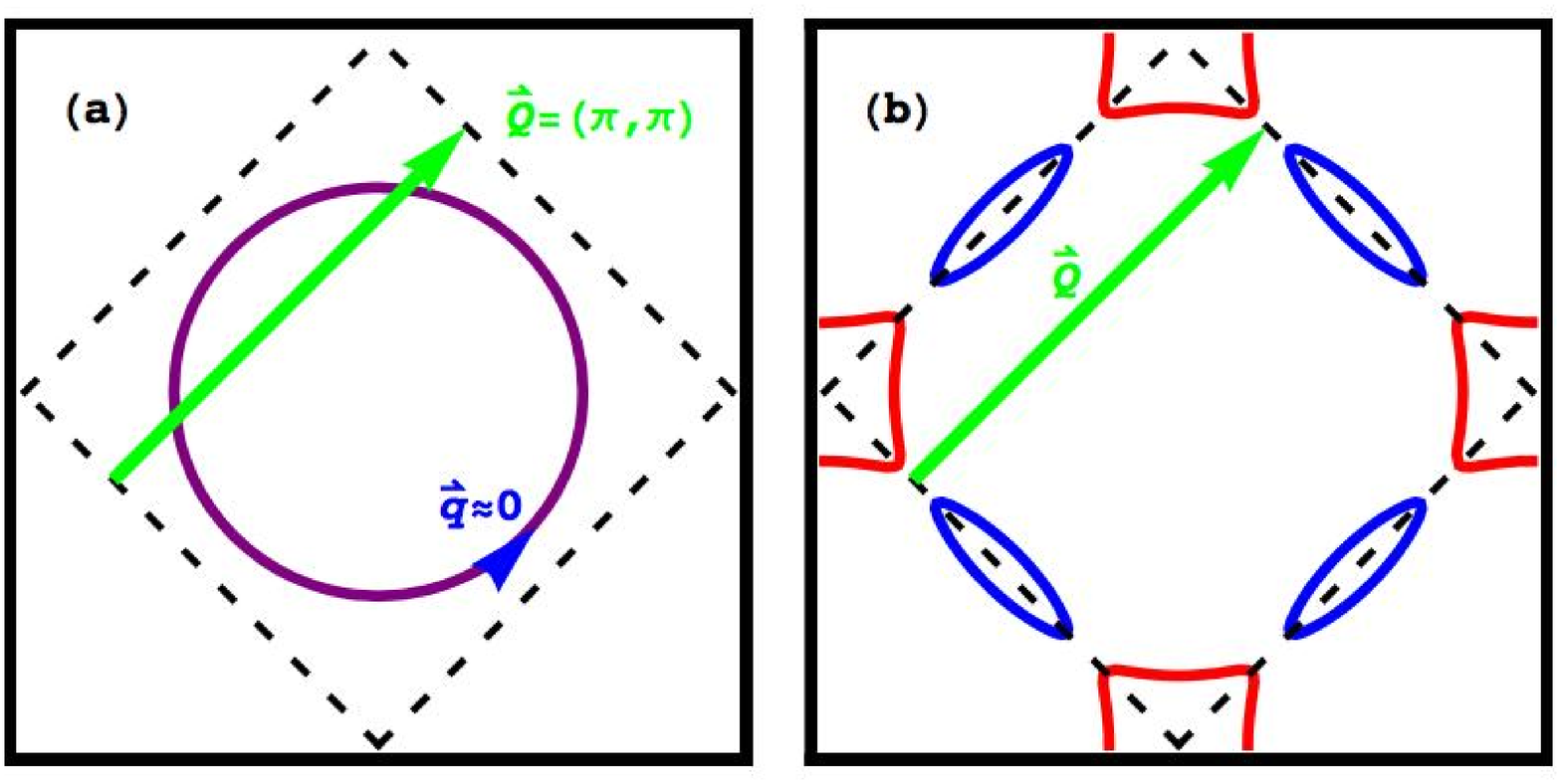}
   \caption[Fermi surface and the ordering wavevector.]
   {
(a) When the Fermi surface of the conduction electrons does not 
intersect the boundary of the antiferromagnetic Brillouin zone 
(AFBZ, the dashed lines), the only 
component of the 
Kondo coupling to the QNL$\sigma $M fields involves forward scattering
for the conduction electrons near the Fermi surface.
(b) (from Ref.~\cite{Yamamoto08}) When the Fermi surface 
of the conduction electrons does 
{not} 
intersect the AFBZ, 
the Kondo coupling connects the re-diagonalized
fermions (whose Fermi surfaces are given by the solid lines)
to the QNL$\sigma $M fields. This coupling is also forward scattering 
in the AFBZ.
}
\label{fig1}
\end{figure}

We expand around the limit $J_K=0$, where the local-moment and 
conduction-electron components are decoupled.
We will consider, for simplicity, square or cubic lattices, although
our results will be generally valid provided the ground 
state
is a
collinear antiferromagnet.
$\mathcal{H}_f$ can be mapped to a 
%quantum non-linear sigma model (QNL$\sigma$M) 
QNL$\sigma$M 
by standard means \cite{Haldane1983, Chakravarty1989}.
The low-lying excitations are concentrated in momentum space near
${\vec q}={\vec Q}$ (the staggered magnetization) and 
near ${\vec q}={\vec 0}$ 
(the total magnetization being conserved):
\begin{equation}
	2 \vec{S}_i 
\to \eta_{\vec{x}} \vec{n}(\vec{x},\tau)\sqrt{1-\left( 
2a^d \vec{L}(\vec{x},\tau) \right)^2} 
+ 
2a^d
\vec{L}(\vec{x},\tau) 
\end{equation}
where $\vec{x}$ labels the position,
$\eta_{\vec{x}} = \pm 1$ on even and odd sites,
$a$ is the lattice constant,
and we have used $S=1/2$.
The linear coupling $\vec{n}\cdot\vec{s}_c $ cannot connect two 
points on the Fermi surface and is hence unimportant for low energy
physics (such a kinematic constraint has appeared in other
contexts, {\it e.g.} Ref.~\cite{Sachdev1995});
see Fig. \ref{fig1}a. 
The Kondo coupling is then replaced by an effective one,
$\vec{S} \cdot\vec{s}_c \rightarrow 
a^d
\vec{L}\cdot\vec{s}_c$,
corresponding to forward scattering for the conduction electrons;
see Fig. \ref{fig1}a. 

%%%%%%%%%%%

The mapping to the QNL$\sigma$M can now be implemented by
integrating out the $\vec{L}$
field. The effective action 
is
\begin{eqnarray}
	\mathcal{S} &=& \mathcal{S}_{\text{QNL}\sigma\text{M}}+\mathcal{S}_{\text{Berry}}+\mathcal{S}_K+\mathcal{S}_c\\
	\mathcal{S}_{\text{QNL}\sigma\text{M}} &\equiv& \frac{c}{2g}\int d^dxd\tau\left[ \left(\nabla \vec{n}(\vec{x},\tau)\right)^2 + 
\left(\frac{\partial\vec{n}(\vec{x},\tau)}{ c ~\partial\tau}\right)^2 \right] 
\nonumber \\
	\mathcal{S}_K &\equiv& 
\lambda\int d^dx d\tau\left[ \vec{s}_c(\vec{x},\tau)\cdot \vec{\varphi}(\vec{x},\tau) \right] \nonumber \\
	\mathcal{S}_c &\equiv& \int d^d K d\varepsilon 
\sum_{\sigma} \psi^{\dagger}_{\sigma}(\vec{K},i\varepsilon) 
\left(i\varepsilon - \xi_{K} \right)\psi_{\sigma}(\vec{K},i\varepsilon)  + \lambda^2\int \psi^4
\nonumber
\end{eqnarray}
where $\xi_K \equiv v_F(K-K_F)$.
The Berry phase term for the
$\vec{n}$ field,
$\mathcal{S}_{\text{Berry}}$, 
is not important
inside the N\'{e}el phase,
which is very different from the ferromagnetic case.
What is meant by ``the Berry phase''
requires some clarification.  Certainly some aspects
of the geometric term do indeed contribute to the physics, but
this will be spelled out in the next section.
We have introduced a vector boson field $\vec{\varphi}$ which 
is shorthand for
$\vec{n}\times\frac{\partial\vec{n}}{\partial\tau}$.  
The $\vec{n}$ field satisfies the constraint $\vec{n}^2=1$, which is solved by
$\vec{n} = (\vec{\pi}, \sigma)$, where 
$\vec{\pi}$ labels the Goldstone magnons
and $\sigma \equiv \sqrt{1-\vec{\pi}^2}$ is the massive field.
We will consider the case of a
spherical Fermi surface; since only forward scattering is important,
our results will apply 
for more complicated Fermi-surface geometries.
The parameters for the QNL$\sigma$M 
will be considered as phenomenological \cite{Chakravarty1989},
though they can be explicitly written in terms of the 
microscopic parameters.
The effective Kondo coupling is
$\lambda = iJ_K/(4dIa^d)$, which will be explicitly demonstrated below.

This summarizes the structure and setup of the effective field theory
for the antiferromagnetic phase of the Kondo lattice model.
We now describe the details of how this is done.

%%%%%%%%%%%%
\subsection{Coherent state representation of the partition function}
%%%%%%%%%%%%

We set up our notation by considering in some detail the 
standard case of the Heisenberg model.  This will help us perform the analogous
mapping for the Kondo lattice model, which is essentially the same but includes
conduction electron coupling.
We will focus on a square lattice with nearest-neighbor (nn) and 
next-nearest-neighbor (nnn) spin-exchange interactions, $I_1$ and $I_2$ respectively. 
%The procedure and 
%the results are essentially the same for other bipartite lattices.
\begin{eqnarray}
%	H 	&=& \sum_{i,j}I_{ij} \mathbf{S}_i\cdot \mathbf{S}_j \nonumber\\
%		&=& I_1 \sum_{\langle ij\rangle}^{nn}\mathbf{S}_i\cdot \mathbf{S}_j 
%+ I_2\sum_{\langle\langle ij\rangle\rangle}^{nnn}\mathbf{S}_i\cdot \mathbf{S}_j \nonumber\\		
%		&=& I_1 \sum_{x,\alpha}\mathbf{S}_x\cdot \mathbf{S}_{x+\alpha} + I_2\sum_{x,\beta}\mathbf{S}_x\cdot \mathbf{S}_{x+\beta}
H 	= \sum_{i,j}I_{ij} \mathbf{S}_i\cdot \mathbf{S}_j 
= I_1 \sum_{\langle ij\rangle}^{nn}\mathbf{S}_i\cdot \mathbf{S}_j 
+ I_2\sum_{\langle\langle ij\rangle\rangle}^{nnn}\mathbf{S}_i\cdot 
\mathbf{S}_j 
= I_1 \sum_{x,\alpha}\mathbf{S}_x\cdot \mathbf{S}_{x+\alpha} + I_2\sum_{x,\beta}\mathbf{S}_x\cdot \mathbf{S}_{x+\beta}
\nonumber\\
\end{eqnarray}
where $x$ runs over all lattice sites, $\alpha$ runs over the nn sites 
for each lattice site $x$, and $\beta$ runs over
nnn sites for each site $x$. 
After the mapping is completed, it will be clear that the 
coupling constant $g$ of the 
%quantum non-linear sigma model (QNL$\sigma$M)
QNL$\sigma$M
can be tuned by changing $I_1$ and $I_2$.

The coherent state spin path integral representation of the partition function for quantum spin systems is now a standard
formalism that can be found in textbooks (e.g. \cite{Fradkin1998, Nagaosa1999}).  We will briefly sketch the main idea.  The partition
function can be written
\begin{eqnarray}
	Z  	&=& \text{Tr} \;e^{-\beta H}
	=
\lim_{M\rightarrow \infty}\int \mathcal{D}\bm{\Lambda}(\tau) \prod_{k=1}^{M-1} \langle \bm{\Lambda}
(\tau_{k+1})\rvert 1-\epsilon H \lvert \bm{\Lambda}(\tau_{k})  \rangle
\end{eqnarray}
where the many-particle
basis is a direct product of single-site spin coherent states: 
$\lvert \bm{\Lambda}(\tau)  \rangle \equiv \prod_x \lvert \bm{\Omega}_x(\tau)\rangle$.  
Here $M$ is the number of discrete time slices in the Trotter decomposition,
and $\epsilon=\beta/M$. Recall that at each lattice site, 
%$\langle \bm{\Omega}\rvert \mathbf{S} \lvert \bm{\Omega}^{\prime}\rangle 
%= S\bm{\Omega}\langle \bm{\Omega}\rvert \bm{\Omega}^{\prime}\rangle$, 
%or
$\langle \bm{\Omega}\rvert \mathbf{S} \lvert \bm{\Omega}\rangle 
= S\bm{\Omega}$,
where the unit spin vector is represented by
\begin{equation}
	\bm{\Omega} = \frac{1}{\sqrt{S(S+1)}}\mathbf{S} 
\approx \mathbf{S}/S = (\cos\phi\sin\theta, \sin\phi\sin\theta, \cos\theta). \nonumber
\end{equation}
Using this (overcomplete) basis, the matrix elements of the Heisenberg Hamiltonian can thus be written,
to leading order in $\epsilon$,
\begin{eqnarray}
	\langle \bm{\Lambda}(\tau_{k+1})\rvert H \lvert \bm{\Lambda}(\tau_{k})  \rangle 
		 &=& S^2 \langle \bm{\Lambda}(\tau_{k+1})\rvert \bm{\Lambda}(\tau_{k})  \rangle \nonumber\\
		 &&\times
\left(  I_1 \sum_{x,\alpha}\bm{\Omega}_x(\tau_k)\cdot \bm{\Omega}_{x+\alpha}(\tau_k) 
+ I_2\sum_{x,\beta}\bm{\Omega}_x(\tau_k)\cdot \bm{\Omega}_{x+\beta}(\tau_k) \right) \nonumber\\
\end{eqnarray}
This allows us to write the Hamiltonian in terms of classical variables $\bm{\Omega}_x$. To linear order in $\epsilon$,
\begin{equation}
	 \langle \bm{\Lambda}(\tau_{k+1})\rvert (1-\epsilon H) \lvert \bm{\Lambda}(\tau_{k})  \rangle
		= \langle \bm{\Lambda}(\tau_{k+1})\rvert  \bm{\Lambda}(\tau_{k}) \rangle e^{-\epsilon H_{cl}(\tau_k)}
\end{equation}
where
\begin{equation}
	H_{cl}(\tau) \equiv S^2I_1 \sum_{x,\alpha}\bm{\Omega}_x(\tau)\cdot \bm{\Omega}_{x+\alpha}(\tau) 
+ S^2I_2\sum_{x,\beta}\bm{\Omega}_x(\tau)\cdot \bm{\Omega}_{x+\beta}(\tau)
\end{equation}
The penalty for the classical representation is the additional overlap $ \langle \bm{\Lambda}(\tau_{k+1})\rvert 
\bm{\Lambda}(\tau_{k})  
\rangle$ which is the Berry phase accumulated from the adiabatic evolution from the time-slice $\tau_k$ to $\tau_{k+1}$. 
Including the Berry
phase accounts for quantum corrections,
and is crucial to obtain the proper mapping to the QNL$\sigma$M.
%%The topological Hopf invariant is only part of the above Berry phase.
We can write it more clearly as follows:
\begin{eqnarray}
	\langle\bm{\Lambda}(\tau_{k+1})\rvert\bm{\Lambda}(\tau_{k})\rangle
	&=& \prod_{x,x^{\prime}} \langle\bm{\Omega}_x(\tau_{k+1})\rvert \bm{\Omega}_{x^{\prime}}(\tau_{k})\rangle
\nonumber \\
	&=& \prod_x \langle\bm{\Omega}_x(\tau_{k+1})\rvert \bm{\Omega}_x(\tau_{k})\rangle \nonumber \\
	&=& \prod_x e^{-iS[1-\cos\theta_x(\tau_{k}) ][\phi_x(\tau_{k+1})-\phi_x(\tau_k)]}
\end{eqnarray}
In the partition function we need an infinite product of such overlaps, which leads to a continuum representation in imaginary time
\begin{eqnarray}
	\lim_{M\rightarrow \infty} \prod_{k=1}^{M-1}\langle\bm{\Lambda}(\tau_{k+1})\rvert\bm{\Lambda}(\tau_{k})\rangle
	&=& \lim_{M\rightarrow \infty}  \prod_{x,k} e^{-iS[1-\cos\theta_x(\tau_{k})][\phi_x(\tau_{k+1})-\phi_x(\tau_k)]}
\nonumber\\
	&=& \lim_{M\rightarrow \infty}  e^{-iS\sum_{x,k}[1-\cos\theta_x(\tau_{k})][\phi_x(\tau_{k+1})-\phi_x(\tau_k)]}
\nonumber\\
	&=& e^{-iS\sum_{x}\int_0^{\beta}d\tau [1-\cos\theta_x(\tau)]\frac{d\phi_x}{d\tau}}
\nonumber\\
	&=& e^{-iS\sum_x\omega(\bm{\Omega}_x)}
\end{eqnarray}
where $\omega(\bm{\Omega}_x)=\int_0^{\beta}d\tau [1-\cos\theta_x(\tau)]\frac{d\phi_x}{d\tau}$ is the Berry phase for a single spin
at site $x$.  
%The important thing to 
Note that the total Berry phase contribution to the action in the path 
integral is given by the sum of the 
Berry phases of all the lattice site spins: $\mathcal{S}_B 
= iS\sum_x\omega(\bm{\Omega}_x)$.  
We have represented the Berry phase of a single spin 
with a set of parameters $\theta_x$ and $\phi_x$ for familiarity,
but we will work with an alternative and convenient representation
given by
\begin{eqnarray}
	\omega(\bm{\Omega}) &=& \int_0^{\beta}d\tau\int_0^1du\Bigg[ \bm{\Omega}(\tau,u)\cdot 
\frac{\partial\bm{\Omega}(\tau,u)}{\partial u}\times\frac{\partial\bm{\Omega}(\tau,u)}{\partial\tau}\Bigg]
\end{eqnarray}
where by convention $\bm{\Omega}(\tau, u=1)=\bm{\Omega}(\tau)$ and $\bm{\Omega}(\tau, u=0)
=(0,0,1)=\lvert 0 \rangle = \lvert S, m=S \rangle$ (see, for example, Ref.~\cite{Sachdev1999}).  
$\bm{\Omega}(\tau,u)$ is a homotopically equivalent continuous deformation of $\bm{\Omega}(\tau)$.

Now, in a similar way to what was done above for the Berry phase, we can take the continuum limit to express
the Hamiltonian term, $H_{cl}$, as an integration over imaginary time.  The partition function then becomes
\begin{eqnarray}
	Z  			&=&\int \mathcal{D}\bm{\Lambda}(\tau)e^{-\mathcal{S}}\\
	\mathcal{S} 	&=& \mathcal{S}_B + \int_0^{\beta}d\tau H_{cl}(\tau)\\
	\mathcal{S}_B 	&=& iS\sum_x\int_0^{\beta}d\tau\int_0^1du\Bigg[ \bm{\Omega}_x(\tau,u)\cdot 
\frac{\partial\bm{\Omega}_x(\tau,u)}{\partial u}\times
\frac{\partial\bm{\Omega}_x(\tau,u)}{\partial\tau}\Bigg]\label{eq:berryPhase}\\
	H_{cl}(\tau)	&=& S^2I_1 \sum_{x,\alpha}^{nn}\bm{\Omega}_x(\tau)\cdot 
\bm{\Omega}_{x+\alpha}(\tau) + S^2I_2\sum_{x,\beta}^{nnn}\bm{\Omega}_x(\tau)\cdot \bm{\Omega}_{x+\beta}(\tau)\label{eq:Hamiltonian}
\end{eqnarray}

%%%%%%%%%%%%%%%%%
\subsection{QNL$\sigma$M mapping for the Heisenberg model}
%%%%%%%%%%%%

We assume an antiferromagnetic order in which case we can represent each spin as the sum of a staggered component $\mathbf{n}_x$
representing the local N\'eel field, and uniform ($q \approx 0$)
fluctuations $\mathbf{L}_x$,
\begin{eqnarray}\label{eq:spinDecomposition}
	\bm{\Omega}_x(\tau) &\equiv&  \eta_x \mathbf{n}_x(\tau) \sqrt{1-\left(\frac{a^d}{S}\mathbf{L}_x(\tau)\right)^2} 
+ \frac{a^d}{S}\mathbf{L}_x(\tau)
\end{eqnarray}
The factor $ \eta_x \in \pm1$ 
%is either positive or negative, 
depending on which sublattice $x$ falls in, while the factor $a^d/S$ 
in front of the uniform
fluctuation field $\mathbf{L}$ ensures that integrating around any small volume will yield the total magnetization contained
in that volume.  
Here $a$ is the lattice constant.
Recall that the spin variable is constrained by the condition $\bm{\Omega}_x\cdot\bm{\Omega}_x=1$ at each site.
With the above choice, this constraint now becomes $\mathbf{n}_x\cdot\mathbf{n}_x = 1$ and $\mathbf{n}_x \cdot \mathbf{L}_x = 0$.
Note that the total number of degrees
of freedom in the system remains the same because in the $\mathbf{n}$, $\mathbf{L}$ representation we must restrict ourselves
to the antiferromagnetic 
Brillouin zone (AFBZ); there are twice as many degrees of freedom 
on half as many sites.

We now wish to write the action in terms of $\mathbf{n}$ and $\mathbf{L}$ rather than $\bm{\Omega}$.  We will consider the Berry phase first,
then the $H_{cl}$ term.

%%%%%%%%%%%%%
\subsection{Berry phase}
%%%%%%%%%%%%
Let us first consider what happens when we substitute this expression for the spin into the Berry phase part of the action.  
We need the expressions for the $\tau$ and $u$ derivatives:
\begin{eqnarray}
	\frac{\partial\bm{\Omega}_x}{\partial u} &\equiv& \bm{\Omega}_u= \eta\gamma\mathbf{n}_u- 
\frac{\eta}{\gamma}\frac{a^d}{S}(\mathbf{L}\cdot\mathbf{L}_u)\mathbf{n} + \frac{a^d}{S}\mathbf{L}_u\\
	\frac{\partial\bm{\Omega}_x}{\partial \tau} &\equiv& \bm{\Omega}_{\tau}= \eta\gamma\mathbf{n}_{\tau}- 
\frac{\eta}{\gamma}\frac{a^d}{S}(\mathbf{L}\cdot\mathbf{L}_{\tau})\mathbf{n} + \frac{a^d}{S}\mathbf{L}_{\tau}
\end{eqnarray}
where we temporarily dropped the site index, $x$, and instead use subscripts to denote differentiation.  We have 
also defined $\gamma \equiv \sqrt{1-\left(\frac{a^d}{S}\mathbf{L}\right)^2}$.  Plugging this into equation (\ref{eq:berryPhase}):
\begin{eqnarray}
\mathcal{S}_B &=& iS\sum_x\int_0^{\beta}d\tau\int_0^1du\Bigg[ \left(\eta\gamma\mathbf{n}+\frac{a^d}{S}\mathbf{L} \right)\cdot
\Big( \eta^2\gamma^2\mathbf{n}_u\times\mathbf{n}_{\tau} - \eta^2\frac{a^d}{S}(\mathbf{L}\cdot\mathbf{L}_u)\mathbf{n}_u
\times\mathbf{n} \nonumber \\
	&\quad&+\eta\gamma\frac{a^d}{S}\mathbf{n}_u\times\mathbf{L}_{\tau}  %\nonumber \\
%	&\quad& 
	- \eta^2\frac{a^d}{S}(\mathbf{L}\cdot\mathbf{L}_u)\mathbf{n}\times\mathbf{n}_{\tau}
+\frac{\eta^2a^{2d}}{\gamma^2S^2}(\mathbf{L}\cdot\mathbf{L}_u)(\mathbf{L}\cdot\mathbf{L}_{\tau})\mathbf{n}\times\mathbf{n}
 \nonumber \\
	&\quad&
	-\frac{\eta}{\gamma}\frac{a^{2d}}{S^2}(\mathbf{L}\cdot\mathbf{L}_u)\mathbf{n}\times\mathbf{L}_{\tau}
	+\eta\gamma\frac{a^d}{S}\mathbf{L}_u\times\mathbf{n}_{\tau}-\frac{\eta}{\gamma}
\frac{a^{2d}}{S^2}(\mathbf{L}\cdot\mathbf{L}_{\tau})\mathbf{L}_u\times\mathbf{n} 
+ \frac{a^{2d}}{S^2}\mathbf{L}_u\times\mathbf{L}_{\tau}  \Big)\Bigg] \nonumber \\
\end{eqnarray}
To simplify this equation requires knowing that $\mathbf{n}_u$, $\mathbf{n}_{\tau}$ and $\mathbf{L}$ are all perpendicular to $\mathbf{n}$. 
This means their triple product must vanish: $\mathbf{L}\cdot\mathbf{n}_u\times\mathbf{n}_{\tau}=0$. 
We also neglect terms higher than linear order in 
%%the spin fluctuation field 
$\mathbf{L}$
[terms quadratic in $\mathbf{L}$ are small compared to those kept in Eq.~(\ref{eq:discreteHamiltonian})], leading to
\begin{eqnarray}
\mathcal{S}_B &\approx& iS\sum_x\int_0^{\beta}d\tau\int_0^1du\Bigg[\eta^3\gamma^3\mathbf{n}\cdot\mathbf{n}_u
\times\mathbf{n}_{\tau}+\eta^2\gamma^2\frac{a^d}{S}\mathbf{n}\cdot(\mathbf{n}_u\times\mathbf{L}_{\tau}+\mathbf{L}_u\times\mathbf{n}_{\tau})\Bigg] \nonumber \\
\end{eqnarray}
%%To be consistent with only keeping terms up to linear order in $\mathbf{L}$, we should set $\gamma \approx 1$.  Also note 
Note also 
that $ \eta_x^2=1$ and $ \eta_x^3= \eta_x$.  We then obtain
\begin{eqnarray}
\mathcal{S}_B 
	&\approx& iS\sum_x\int_0^{\beta}d\tau\int_0^1du\Bigg[ \eta\mathbf{n}\cdot\mathbf{n}_u\times\mathbf{n}_{\tau}
+\frac{a^d}{S}\mathbf{n}\cdot(\mathbf{n}_u\times\mathbf{L}_{\tau}+\mathbf{L}_u\times\mathbf{n}_{\tau})\Bigg] 
\nonumber \\
	&=& iS\sum_x\int_0^{\beta}d\tau\int_0^1du\Bigg[ \eta_x\mathbf{n}_x\cdot\frac{\partial\mathbf{n}_x}
{\partial u}\times\frac{\partial\mathbf{n}_x}{\partial\tau}+\frac{a^d}{S}\mathbf{n}_x\cdot
\Bigg( \frac{\partial\mathbf{n}_x}{\partial u}\times\frac{\partial\mathbf{L}_x}{\partial\tau} %\nonumber\\
%&&
+ \frac{\partial\mathbf{L}_x}{\partial u}\times\frac{\partial\mathbf{n}_x}{\partial\tau}  \Bigg)\Bigg]
\nonumber\\
	&=& iS\sum_x\int_0^{\beta}d\tau\int_0^1du\Bigg[ \eta_x\mathbf{n}_x\cdot\frac{\partial\mathbf{n}_x}
{\partial u}\times\frac{\partial\mathbf{n}_x}{\partial\tau}+\frac{a^d}{S}\frac{\partial}{\partial\tau}
\left(\mathbf{n}_x\cdot\frac{\partial\mathbf{n}_x}{\partial u}\times\mathbf{L}_x\right) \nonumber\\
&&+ \frac{a^d}{S}\frac{\partial}{\partial u}\left( \mathbf{n}_x\cdot\mathbf{L}_x\times\frac{\partial\mathbf{n}_x}{\partial\tau} \right)\Bigg]
\end{eqnarray}
In the second line we have restored the full notation, while the third line can be written with total derivatives since the terms
proportional to $\frac{\partial^2\mathbf{n}}{\partial\tau\partial u}$ cancel thanks to the triple product identity 
$\zeta\cdot\mathbf{b}\times\mathbf{c} = -\mathbf{b}\cdot\zeta\times\mathbf{c}$.  The second term in the third 
line vanishes after integrating the total $\tau$ derivative and using the periodicity of the fields.  The third term in the third 
line can be integrated over $u$, and the value at $u=0$ is zero due to the orthogonality at the north pole.  We finally find,
\begin{eqnarray}
\mathcal{S}_B &=& iS\sum_x \eta_x\int_0^{\beta}d\tau\int_0^1du\left(\mathbf{n}_x\cdot\frac{\partial\mathbf{n}_x}{\partial u}
\times\frac{\partial\mathbf{n}_x}{\partial\tau}\right) - i\sum_x a^d\int_0^{\beta}d\tau\left( \mathbf{L}_x\cdot\mathbf{n}_x
\times\frac{\partial\mathbf{n}_x}{\partial\tau} \right) \nonumber \\
\end{eqnarray}
The first term is precisely the Berry phase for the N\'eel component $\mathbf{n}$, while the second term is something additional
that must be added to the total action.  Although both terms came from the expression for the Berry phase of $\bm{\Omega}$,
it is only the first term that is often referred to as the Berry phase for the antiferromagnet.

%%%%%%%%%%%%
\subsection{Hamiltonian}
%%%%%%%%%%%%
Next we compute the contribution to the action from the Hamiltonian $H_{cl}(\tau)$ expressed in terms of $\mathbf{n}$ and 
$\mathbf{L}$ fields.  Plugging (\ref{eq:spinDecomposition}) into (\ref{eq:Hamiltonian}),
\begin{eqnarray}
	H_{cl}(\tau)
	&=& S^2I_1 \sum_{x,\alpha}^{nn} \Bigg[  \eta_x \mathbf{n}_x(\tau) \sqrt{1-\left(\frac{a^d}{S}\mathbf{L}_x(\tau)\right)^2} 
+ \frac{a^d}{S}\mathbf{L}_x(\tau)\Bigg] \nonumber \\
	&&\cdot \Bigg[\eta_{x+\alpha} \mathbf{n}_{x+\alpha}(\tau) 
\sqrt{1-\left(\frac{a^d}{S}\mathbf{L}_{x+\alpha}(\tau)\right)^2} + \frac{a^d}{S}\mathbf{L}_{x+\alpha}(\tau)\Bigg] \nonumber \\
	&\quad& + S^2I_2\sum_{x,\beta}^{nnn}\Bigg[  \eta_x \mathbf{n}_x(\tau) \sqrt{1-\left(\frac{a^d}{S}\mathbf{L}_x(\tau)\right)^2} 
+ \frac{a^d}{S}\mathbf{L}_x(\tau)\Bigg] \nonumber \\
	&&\cdot \Bigg[\eta_{x+\beta} \mathbf{n}_{x+\beta}(\tau) 
\sqrt{1-\left(\frac{a^d}{S}\mathbf{L}_{x+\beta}(\tau)\right)^2} + \frac{a^d}{S}\mathbf{L}_{x+\beta}(\tau)\Bigg] \nonumber\\
	&\approx& S^2I_1 \sum_{x,\alpha}^{nn} \Bigg[  \eta_x \mathbf{n}_x(\tau) \left(1-\frac{1}{2}\left(\frac{a^d}{S}\mathbf{L}_x(\tau)\right)^2\right) 
+ \frac{a^d}{S}\mathbf{L}_x(\tau)\Bigg] \nonumber \\
	&&\cdot \Bigg[\eta_{x+\alpha} \mathbf{n}_{x+\alpha}(\tau) 
\left(1-\frac{1}{2}\left(\frac{a^d}{S}\mathbf{L}_{x+\alpha}(\tau)\right)^2\right) + \frac{a^d}{S}\mathbf{L}_{x+\alpha}(\tau)\Bigg] \nonumber \\
	&\quad& + S^2I_2\sum_{x,\beta}^{nnn}\Bigg[  \eta_x \mathbf{n}_x(\tau) \left(1-\frac{1}{2}\left(\frac{a^d}{S}\mathbf{L}_x(\tau)\right)^2\right) 
+ \frac{a^d}{S}\mathbf{L}_x(\tau)\Bigg] \nonumber \\
	&&\cdot \Bigg[\eta_{x+\beta} \mathbf{n}_{x+\beta}(\tau) 
\left(1-\frac{1}{2}\left(\frac{a^d}{S}\mathbf{L}_{x+\beta}(\tau)\right)^2\right) + \frac{a^d}{S}\mathbf{L}_{x+\beta}(\tau)\Bigg] \nonumber\\
\end{eqnarray}
Since $\mathbf{n}$ is a unit vector, we have the identity $\mathbf{n}_x\cdot\mathbf{n}_y = 1-\frac{1}{2}(\mathbf{n}_x-\mathbf{n}_y)
\cdot(\mathbf{n}_x-\mathbf{n}_y)$.  We also know that at every site $\mathbf{n}_x\cdot\mathbf{L}_x = 0$.  Using these two identities
and dropping the $\tau$ label for brevity, the Hamiltonian can be expressed as follows:
\begin{eqnarray}
	H_{cl} 
		&=& S^2 I_1\sum_{x,\alpha}\Bigg\{ \eta_{x}\eta_{x+\alpha}\left[ 1-\frac{1}{2}(\mathbf{n}_x-\mathbf{n}_y)
\cdot(\mathbf{n}_x-\mathbf{n}_y) \right]  \left[ 1- \frac{ a^{2d} }{ 2S^2 }( \mathbf{L}_x^2 + \mathbf{L}_{x+\alpha}^2 ) \right] \nonumber\\
		&&+ \frac{a^d}{2S^2}\eta_x\mathbf{n}_{x}\cdot(\mathbf{L}_{x+\alpha}-\mathbf{L}_x) + \frac{a^d}{2S^2}\eta_{x+\alpha}
\mathbf{n}_{x+\alpha}\cdot(\mathbf{L}_{x}-\mathbf{L}_{x+\alpha}) \nonumber \\
		&& + \frac{a^{2d} }{ 2S^2}\left[\mathbf{L}_{x}^2+\mathbf{L}_{x+\alpha}^2
-(\mathbf{L}_x-\mathbf{L}_{x+\alpha} )^2\right]
			\Bigg\}\nonumber\\
		&&+S^2 I_2\sum_{x,\beta}\Bigg\{ \eta_{x}\eta_{x+\beta}\left[ 1-\frac{1}{2}(\mathbf{n}_x-\mathbf{n}_y)
\cdot(\mathbf{n}_x-\mathbf{n}_y) \right]  \left[ 1- \frac{ a^{2d} }{ 2S^2 }( \mathbf{L}_x^2 + \mathbf{L}_{x+\beta}^2 ) \right] \nonumber\\
		&&+ \frac{a^d}{2S^2}\eta_x\mathbf{n}_{x}\cdot(\mathbf{L}_{x+\beta}-\mathbf{L}_x) 
+ \frac{a^d}{2S^2}\eta_{x+\beta}\mathbf{n}_{x+\beta}\cdot(\mathbf{L}_{x}-\mathbf{L}_{x+\beta}) \nonumber \\
		&&+ \frac{a^{2d} }{ 2S^2}\left[\mathbf{L}_{x}^2+\mathbf{L}_{x+\beta}^2-(\mathbf{L}_x-\mathbf{L}_{x+\beta} )^2\right]
			\Bigg\}
\end{eqnarray}
where we used $2\mathbf{L}_x\cdot\mathbf{L}_{x+\alpha} = \mathbf{L}_{x}^2+\mathbf{L}_{x+\alpha}^2-(\mathbf{L}_x
-\mathbf{L}_{x+\alpha} )^2$ and similarly for $\beta$.  We have written the expression in this way in order to take advantage
of a Taylor series expansion between different lattice sites: $n_{x+\alpha}^{b}-n_{x}^{b} \approx a(\vec{\alpha}\cdot\nabla)n_x^{b} + \cdots$ and $n_{x+\beta}^{b}-n_{x}^{b} \approx a\sqrt{2}(\vec{\beta}\cdot\nabla)n_x^{b} + \cdots$.
%\begin{eqnarray}
%	n_{x+\alpha}^{b}-n_{x}^{b} &\approx& a(\vec{\alpha}\cdot\nabla)n_x^{b} + \cdots \\
%	n_{x+\beta}^{b}-n_{x}^{b} &\approx& a\sqrt{2}(\vec{\beta}\cdot\nabla)n_x^{b} + \cdots
%\end{eqnarray}
The index $b$ runs over the components of the vector field.  Note that nearest neighbor (nn) sites $x$ and $x+\alpha$ are separated by 
a distance $a$, while next nearest neighbor (nnn) sites $x$ and $x+\beta$ are separated by a distance $a\sqrt{2}$.  Expressing all lattice
differences in this way leads to:
\begin{eqnarray}
	H_{cl} 
		&=& S^2 I_1\sum_{x,\alpha}\Bigg\{ \eta_{x}\eta_{x+\alpha}\left[ 1-\frac{a^2}{2}[(\vec{\alpha}\cdot\nabla)n_x^{b}]^2 \right] \nonumber\\
		&&+ \frac{a^{2d}}{2S^2}\left[ (1-\eta_{x}\eta_{x+\alpha})(\mathbf{L}_x^2+\mathbf{L}_{x+\alpha}^2)
-a^2[(\vec{\alpha}\cdot\nabla)L_x^{b}]^2\right] \nonumber\\
		&&+ \frac{a^{d+1}}{S^2}\left[\eta_x n_x^b\nabla L_x^b-\eta_{x+\alpha} n_{x+\alpha}^b\nabla L_{x+\alpha}^b\right]
			\Bigg\}\nonumber\\
		&&+ S^2 I_2\sum_{x,\beta}\Bigg\{ \eta_{x}\eta_{x+\beta}\left[ 1-\frac{2a^2}{2}[(\vec{\beta}\cdot\nabla)n_x^{b}]^2\right] \nonumber\\
		&&+ \frac{a^{2d}}{2S^2}\left[ (1-\eta_{x}\eta_{x+\beta})(\mathbf{L}_x^2+\mathbf{L}_{x+\beta}^2)
-2a^2[(\vec{\beta}\cdot\nabla)L_x^{b}]^2\right] \nonumber\\
		&& + \frac{a^{d+1}\sqrt{2} }{ S^2}\left[\eta_x n_x^b\nabla L_x^b-\eta_{x+\beta} n_{x+\beta}^b \nabla L_{x+\beta}^b\right]
			\Bigg\}
\end{eqnarray}
After summing over lattices sites and neighbors, each term proportional to $a^{d+1}$ sums to zero, so,
\begin{eqnarray}
	H_{cl} 
		&=& S^2 I_1\sum_{x,\alpha}\Bigg\{ \eta_{x}\eta_{x+\alpha}\left[ 1-\frac{a^2}{2}[(\vec{\alpha}\cdot\nabla)n_x^{b}]^2 \right] \nonumber\\
	&&+ \frac{a^{2d}}{2S^2}\left[ (1-\eta_{x}\eta_{x+\alpha})(\mathbf{L}_x^2+\mathbf{L}_{x+\alpha}^2) - a^2 [(\vec{\alpha}\cdot\nabla)L_x^{b}]^2 \right] 
			\Bigg\}\nonumber\\
		&&+ S^2 I_2\sum_{x,\beta}\Bigg\{ \eta_{x}\eta_{x+\beta}\left[ 1- a^2[(\vec{\beta}\cdot\nabla)n_x^{b}]^2 \right] \nonumber\\
		&&+ \frac{a^{2d}}{2S^2}\left[ (1-\eta_{x}\eta_{x+\beta})(\mathbf{L}_x^2+\mathbf{L}_{x+\beta}^2) - 2a^2[(\vec{\beta}\cdot\nabla)L_x^{b}]^2 \right]
			\Bigg\}
\end{eqnarray}
Now, to our order of approximation, $\mathbf{L}_{x+\alpha}^2 \approx \mathbf{L}_x^2$.  Also, since $\alpha$ runs over nearest-neighbors, 
while $\beta$ runs over next-nearest-neighbors, we have $ \eta_x\eta_{x+\alpha}=-1$ and $ \eta_x\eta_{x+\beta}=+1$. 
%% So $I_1$ is an inter-sublattice
%%interaction, while $I_2$ is an intra-sublattice interaction.  
On the square lattice, the number of nn and nnn sites is $2d$,
and to avoid double counting we divide by 2.
\begin{eqnarray}
	H_{cl} 
		&=& S^2 I_1d\sum_{x}\Bigg\{ -1 + \frac{a^2}{2d}(\nabla\mathbf{n}_x)^2 + \frac{a^{2d}}{2S^2}\left[ 4\mathbf{L}_x^2 
- \frac{a^2}{d}(\nabla\mathbf{L}_x)^2\right] 
			\Bigg\} \nonumber \\
		&&+ S^2 I_2d\sum_{x}\Bigg\{ 1- \frac{a^2}{d}(\nabla \mathbf{n}_x)^2 + \frac{a^{2d}}{2S^2}\left[ 
- \frac{2a^2}{d}(\nabla\mathbf{L}_x)^2\right]
			\Bigg\}
\end{eqnarray}
Note that $(\nabla \mathbf{n}_x)^2$ should be interpreted as $\sum_{\mu=1}^{d}\sum_{b=1}^{N}\left(\frac{\partial n^b_x}{\partial x^{\mu}}\right)^2$.   
We also used $\sum_{x,\alpha} [(\vec{\alpha}\cdot\nabla)n_x^{b}]^2 = \sum_{x} [(\frac{\partial}{\partial x}+\frac{\partial}{\partial y}
+\frac{\partial}{\partial z})n_x^b]^2 = \sum_x (\nabla\mathbf{n}_x)^2$, and similarly for $\beta$ and $\mathbf{L}$ terms.

To be consistent with our expansion in small powers of $a$ and $1/S$ we should also ignore gradient terms like $(\nabla \mathbf{L})^2$.
The expression simplifies to
\begin{eqnarray}
	H_{cl} 
		&=& S^2 d\sum_{x}\Bigg\{I_2 - I_1 + \frac{a^2}{2d}(I_1-2I_2)(\nabla \mathbf{n}_x)^2+ \frac{2I_1 a^{2d}}{S^2}\mathbf{L}_x^2 \Bigg\} 
\nonumber \\
		&=& S^2 d\mathcal{N}_{site}(I_2 - I_1)+S^2\frac{a^2}{2}(I_1-2I_2)\sum_{x}(\nabla \mathbf{n}_x)^2+ 2dI_1 a^{2d}\sum_x \mathbf{L}_x^2 \label{eq:discreteHamiltonian}
\end{eqnarray}
Finally, we take the continuum limit with the correspondence $\sum_x  \rightarrow a^{-d}\int d^dx$,
\begin{eqnarray}
	H_{cl}(\tau)
		&=& const_1 + S^2a^{2-d}(I_1/2-I_2)\int d^dx\left(\nabla\mathbf{n}(\vec{x},\tau)\right)^2 + 2dI_1 a^{d}\int d^dx \mathbf{L}^2(\vec{x},\tau) \nonumber \\
\label{hamiltonian}
\end{eqnarray}
We have introduced the constant factor $const_1 \equiv \mathcal{N}_{site}dS^2(I_2-I_1)$ which is unimportant for our purposes.

%%%%%%%%%%%%%%%%%%%%%%%%%%%%%%%%%
\subsection{Completing the square and the QNL$\sigma$M mapping}
%%%%%%%%%%%%
At this point the total action is given by
\begin{eqnarray}
	Z &=& \int \mathcal{D}\mathbf{n}\mathcal{D}\mathbf{L}\;\delta(\mathbf{n}^2-1)\delta(\mathbf{n}\cdot\mathbf{L})
e^{-\mathcal{S}[\mathbf{n},\mathbf{L}]} \\
	\mathcal{S}[\mathbf{n},\mathbf{L}]
		&=& const_1 +  \mathcal{S}_{B}^{\prime}[\mathbf{n}] + S^2a^{2-d}(I_1/2-I_2)\int_0^{\beta}d\tau\int d^dx
\left(\nabla\mathbf{n}(\vec{x},\tau)\right)^2 \nonumber\\
		&\quad&+ 2dI_1 a^{d}\int_0^{\beta}d\tau\int d^dx \mathbf{L}^2(\vec{x},\tau)\nonumber\\
		&\quad& - i\int_0^{\beta}d\tau\int d^dx \left( \mathbf{L}(\vec{x},\tau)\cdot\mathbf{n}(\vec{x},\tau)
\times\frac{\partial\mathbf{n}(\vec{x},\tau)}{\partial\tau} \right) \label{eq:totalAction} \\
	\mathcal{S}_{B}^{\prime}[\mathbf{n}]	&\equiv& iS\sum_x \eta_x\int_0^{\beta}d\tau\int_0^1du
\left(\mathbf{n}_x\cdot\frac{\partial\mathbf{n}_x}{\partial u}\times\frac{\partial\mathbf{n}_x}{\partial\tau}\right)
\end{eqnarray}
where the Berry phase $ \mathcal{S}_{B}^{\prime}[\mathbf{n}]$ now only depends on the $\mathbf{n}$ field.
The delta functionals enforce the local constraints.  To deal with them, we use the integral representation 
of the delta functional and introduce a scalar Lagrange multiplier field $\alpha(\vec{x},\tau)$:
\begin{eqnarray}
	\delta(\mathbf{n}\cdot\mathbf{L}) &=& \int \mathcal{D}\alpha \;e^{-\int_0^{\beta}d\tau\int d^dx \; 
i\alpha(\vec{x},\tau)\mathbf{n}(\vec{x},\tau)\cdot\mathbf{L}(\vec{x},\tau)}
\end{eqnarray}

It is now clear that the functional integral is Gaussian with respect to $\mathbf{L}$, so we may ``complete the square''
and integrate it out completely.  If we use the identity $\int\mathcal{D}\mathbf{L}e^{\int-\zeta\mathbf{L}^2
+\mathbf{b}\cdot\mathbf{L}}=\pi^{M/2}(\det\zeta)^{-1/2}e^{\mathbf{b}^2/4\zeta}$, the correspondence 
is $\zeta = 2dI_1a^d$ and $\mathbf{b} = i\mathbf{n}\times\dot{\mathbf{n}}-i\alpha\,\mathbf{n}$.  
We also need the quadruple vector product identity $(\mathbf{n}\times\dot{\mathbf{n}})^2=\dot{\mathbf{n}}^2\mathbf{n}^2
-(\dot{\mathbf{n}}\cdot\mathbf{n})^2$, and the relations $\dot{\mathbf{n}}\cdot\mathbf{n}=0$ and $\mathbf{n}^2=1$ 
and $\mathbf{n}\cdot\mathbf{n}\times\dot{\mathbf{n}}=0$.  This leads to $\frac{\mathbf{b}^2}{4\zeta} = -\frac{1}{8dI_1a^d}
(\dot{\mathbf{n}}^2+\alpha^2)$, and hence
\begin{eqnarray}
	Z &=& \int \mathcal{D}\mathbf{n} \mathcal{D}\alpha\;\delta(\mathbf{n}^2-1) e^{-\mathcal{S}[\mathbf{n},\alpha]} \\
	\mathcal{S}[\mathbf{n},\alpha]
		&=& \beta\, const_1- \frac{1}{2}\log\frac{\pi^M}{2dI_1 a^d} + S^2a^{2-d}(I_1/2-I_2)\int_0^{\beta}d\tau\int d^dx
\left(\nabla\mathbf{n}(\vec{x},\tau)\right)^2 \nonumber \\
		&\quad& + \frac{1}{8dI_1 a^d}\int_0^{\beta}d\tau\int d^dx \left[ \left(\frac{\partial\mathbf{n}(\vec{x},\tau) }
{ \partial\tau}\right)^2 + \alpha^2(\vec{x},\tau) \right] + \mathcal{S}_{B}^{\prime}[\mathbf{n}] 
\nonumber \\
		&\equiv& const_2+\mathcal{S}_{B}^{\prime}[\mathbf{n}] + \frac{c}{2g}\int_0^{\beta}d\tau\int d^dx \Bigg[ 
\left(\nabla\mathbf{n}(\vec{x},\tau)\right)^2 + \frac{1}{c^2}\left(\frac{\partial\mathbf{n}(\vec{x},\tau) }{ \partial\tau}\right)^2 
\Bigg]  \nonumber \\
		&\quad&+\frac{1}{2gc} \int_0^{\beta}d\tau\int d^dx \; \alpha^2(\vec{x},\tau) 
\end{eqnarray}
where we have defined
\begin{eqnarray}
	c &\equiv&	2aSI_1\sqrt{d}\sqrt{\frac{I_1-2I_2}{I_1}}\\
	g &\equiv&	\frac{2a^{d-1}\sqrt{d}}{S}\sqrt{\frac{I_1}{I_1-2I_2}}\\
	const_2 &\equiv&	\beta\, const_1-\frac{1}{2}\log\frac{\pi^M}{2dI_1a^d}
\end{eqnarray}
Recall that $M$ is the number of time slices and may be considered to be of order $\mathcal{N}_{site}$.

All that remains is to perform the gaussian integral over $\alpha$: $\int\mathcal{D}\alpha\;e^{-\frac{1}{2gc}\int \alpha^2} 
= \sqrt{2gc\pi^M}=e^{(1/2)\log(2gc\pi^M)}$.  Our final answer becomes:
\begin{eqnarray}
	Z &=& \int \mathcal{D}\mathbf{n}\;\delta(\mathbf{n}^2-1) e^{-\mathcal{S}[\mathbf{n}]} \\
	\mathcal{S}[\mathbf{n}]
		&=& const_3 + \mathcal{S}_{B}^{\prime}[\mathbf{n}]+\frac{c}{2g}\int_0^{\beta}d\tau\int d^dx \Bigg[ 
\left(\nabla\mathbf{n}(\vec{x},\tau)\right)^2 + \frac{1}{c^2}\left(\frac{\partial\mathbf{n}(\vec{x},\tau) }{ \partial\tau}\right)^2 \Bigg]\nonumber\\
\end{eqnarray}
where $const_3 \equiv const_2 - \frac{1}{2}\log(2gc\pi^M) = \beta\mathcal{N}_{site}S^2 d(I_2-I_1)-\log(2\pi^M)$.  
These results
 for the constants $c$ and $g$ agree with \cite{Einarsson1991} who considered first, second and
third neighbor couplings.  Clearly, by adjusting $I_1$ and $I_2$ we can tune $g$.  This model can also be expressed in terms
of the spin-wave stiffness and transverse magnetic susceptibility.  They are given by
\begin{eqnarray}
	\rho_s &=& \frac{c}{g} = S^2 a^{2-d}(I_1-2I_2)\\
	\chi_{\perp} &=& \frac{1}{cg} = \frac{1}{4a^dI_1d}
\end{eqnarray}
Notice that $\chi_{\perp}$ is independent of $I_2$, which means that nnn interactions do not renormalize the transverse magnetic susceptibility.

%%%%%%%%%%%%%%%
\subsection{QNL$\sigma$M mapping for the Kondo lattice model}
%%%%%%%%%%%%

We have now shown how to map the quantum Heisenberg AF to the QNL$\sigma$M.  Next, we want to incorporate the Kondo interaction
which couples the local moment spin, $\mathbf{S}$, to the conduction election spin, $\mathbf{s}_c$.  This adds the following term:
\begin{eqnarray}
		J_K S\sum_x \bm{\Omega}_x(\tau) \cdot \mathbf{s}_{c,x}(\tau) &=&
		J_K S a^{-d} \int d^dx\,d\tau \;\mathbf{s}_c(\vec{x},\tau) \cdot \bm{\Omega}(\vec{x},\tau) 
\nonumber \\
			&=& J_K S a^{-d} \int d^dx \,d\tau\,  
\Bigg[ \eta_x \big(\mathbf{n}(\vec{x}, \tau)\cdot\mathbf{s}_c(\vec{x},\tau)\big) \nonumber\\
	&&\times 
	\sqrt{1-\left(\frac{a^d}{S}\mathbf{L}(\vec{x}, \tau)\right)^2} %\nonumber \\
%	&&
	+ \frac{a^d}{S}\mathbf{L}(\vec{x}, \tau)\cdot\mathbf{s}_c(\vec{x},\tau) 
\Bigg] \nonumber \\
			&\approx& J_K \int d^dx \,d\tau\,  \mathbf{L}(\vec{x}, \tau)\cdot\mathbf{s}_c(\vec{x},\tau)
\label{eq:AFQNLSMCoupling}
\end{eqnarray}
The last line follows due to $\mathbf{n}\cdot\mathbf{s}_c \approx 0$.  The latter is because we have, as discussed earlier,
chosen to work with a Fermi surface that does not intersect the 
AFBZ boundary (see Fig.~\ref{fig1}a);
what remains of the Kondo interaction is the (nearly) forward scattering channel for the conduction electrons.
The assumption we make is not necessarily that the density of conduction electrons is infinitesimally small,
but only that it does not intersect the 
AFBZ boundary.  Specifically, we require $Q > 2K_F$.
In section \ref{AFBZIntersection} we will discuss the modifications to the theory when the Fermi surface does indeed intersect the magnetic zone boundary~\cite{Yamamoto08}, {\it i.e.} when $Q <2K_F$.

Let us return to the action for the quantum AF before completing the square (equation \ref{eq:totalAction}), and add to that the above Kondo coupling. 
The total action for the Kondo Lattice Model, $\mathcal{S}_{KLM}$, now has something extra coupled to the $\mathbf{L}$ field:
\begin{eqnarray}
	Z &=& \int \mathcal{D}\mathbf{n}\mathcal{D}\mathbf{L}\mathcal{D}\alpha\mathcal{D}\psi^{\dagger}\mathcal{D}\psi\;\delta(\mathbf{n}^2-1) \nonumber\\
	&&\times e^{-\mathcal{S}_{KLM}[\mathbf{n},\mathbf{L},\alpha, \mathbf{s}_c]-\mathcal{S}_c[\psi^{\dagger},\psi]-\mathcal{S}_{B}^{\prime}[\mathbf{n}]} \\
	\mathcal{S}_{KLM}[\mathbf{n},\mathbf{L},\alpha,\mathbf{s}_c] 
		&=& const_1 + S^2a^{2-d}(I_1/2-I_2)\int_0^{\beta}d\tau\int d^dx\left(\nabla\mathbf{n}(\vec{x},\tau)\right)^2 \nonumber\\
		&\quad&+ 2dI_1 a^{d}\int_0^{\beta}d\tau\int d^dx \mathbf{L}^2(\vec{x},\tau) \nonumber \\
		&\quad& - \int_0^{\beta}d\tau\int d^dx \mathbf{L}(\vec{x},\tau)\cdot\Bigg( i\mathbf{n}(\vec{x},\tau)\times
\frac{\partial\mathbf{n}(\vec{x},\tau)}{\partial\tau} - J_K \mathbf{s}_c(\vec{x},\tau) \nonumber \\
		&& - i\alpha(\vec{x},\tau)\mathbf{n}(\vec{x},\tau) \Bigg)
\end{eqnarray}
where $const_1$ and $\mathcal{S}_B^{\prime}[\mathbf{n}]$ are as defined previously, and $\mathcal{S}_c[\psi^{\dagger},\psi]$
is the conduction electron component of the action.  Just like before, the functional integral 
is Gaussian with respect to the $\mathbf{L}$ field.  With the identity $\int\mathcal{D}\mathbf{L}e^{\int-\zeta\mathbf{L}^2
+\mathbf{b}\cdot\mathbf{L}}=\sqrt{\pi^M/\det\zeta}\,e^{\mathbf{b}^2/4\zeta}$, the correspondence is now $\zeta 
= 2dI_1a^d$ and $\mathbf{b} = i\mathbf{n}\times\dot{\mathbf{n}}-J_K \mathbf{s}_c - i\alpha\mathbf{n}$.  The important quantity is:
\begin{eqnarray}
	\frac{\mathbf{b}^2}{4\zeta} &=& \frac{1}{8dI_1 a^d}\Big[-(\mathbf{n}\times\dot{\mathbf{n}})^2 - iJ_K \mathbf{s}_c
\cdot \mathbf{n}\times\dot{\mathbf{n}} + \alpha \mathbf{n}\cdot \mathbf{n}\times\dot{\mathbf{n}}  - iJ_K \mathbf{s}_c\cdot \mathbf{n}\times\dot{\mathbf{n}} +J_K^2\mathbf{s}_c^2\nonumber\\
	&\quad&+iJ_K\alpha\mathbf{s}_c\cdot\mathbf{n} 
	+\alpha\mathbf{n}\cdot\mathbf{n}\times\dot{\mathbf{n}} + iJ_K\mathbf{s}_c\cdot\mathbf{n}
- \alpha^2\mathbf{n}^2
	\Big]
\end{eqnarray}
To simplify this expression, we need the following identities:
$\mathbf{n}^2 = 1$, $\dot{\mathbf{n}}\cdot\mathbf{n} = 0$, $(\mathbf{n}\times\dot{\mathbf{n}})^2 = \dot{\mathbf{n}}^2\mathbf{n}^2-(\dot{\mathbf{n}}\cdot\mathbf{n})^2=-\dot{\mathbf{n}}^2$, $\mathbf{n}\cdot\mathbf{n}\times\dot{\mathbf{n}} = 0$, $\mathbf{s}_c\cdot\mathbf{n} \approx 0$, and
%\begin{eqnarray}
%\mathbf{n}^2 &=&1\\
%\dot{\mathbf{n}}\cdot\mathbf{n}&=&0\\
%(\mathbf{n}\times\dot{\mathbf{n}})^2 &=& \dot{\mathbf{n}}^2\mathbf{n}^2-(\dot{\mathbf{n}}\cdot\mathbf{n})^2=-\dot{\mathbf{n}}^2\\
%\mathbf{n}\cdot\mathbf{n}\times\dot{\mathbf{n}} &=& 0 \\
%\mathbf{s}_c\cdot\mathbf{n} &\approx& 0 \\
%\mathbf{s}_c^2 &=& \sum_{a,\alpha,\beta,\gamma,\delta}\psi^{\dagger}_{\alpha}\frac{\tau^a}{2}
%\psi_{\beta}\psi^{\dagger}_{\gamma}\frac{\tau^a}{2}\psi_{\delta} = \frac{3}{4}\left(\sum_{\sigma}\psi^{\dagger}_{\sigma}\psi_{\sigma} 
%- 2\psi^{\dagger}_{\uparrow}\psi_{\uparrow}\psi^{\dagger}_{\downarrow}\psi_{\downarrow}\right)
%\end{eqnarray}
\begin{eqnarray}
\mathbf{s}_c^2 &=& \sum_{a,\alpha,\beta,\gamma,\delta}\psi^{\dagger}_{\alpha}\frac{\tau^a}{2}
\psi_{\beta}\psi^{\dagger}_{\gamma}\frac{\tau^a}{2}\psi_{\delta} = \frac{3}{4}\left(\sum_{\sigma}\psi^{\dagger}_{\sigma}\psi_{\sigma} 
- 2\psi^{\dagger}_{\uparrow}\psi_{\uparrow}\psi^{\dagger}_{\downarrow}\psi_{\downarrow}\right). \nonumber
\end{eqnarray}
This leads to:
\begin{eqnarray}
	\frac{\mathbf{b}^2}{4\zeta} &=& \frac{1}{8dI_1 a^d}\Big[-\dot{\mathbf{n}}^2 - 2iJ_K \mathbf{s}_c\cdot \mathbf{n}\times\dot{\mathbf{n}} 
 +\frac{3J_K^2 }{4}\left(\sum_{\sigma}\psi^{\dagger}_{\sigma}\psi_{\sigma} - 2\psi^{\dagger}_{\uparrow}\psi_{\uparrow}
\psi^{\dagger}_{\downarrow}\psi_{\downarrow}\right) %\nonumber\\
	%&& 
	- \alpha^2\mathbf{n}^2
	\Big] \nonumber
\end{eqnarray}
Note that the terms that came from $\mathbf{s}_c^2$ serve only to renormalize the direct quadratic and quartic 
fermion couplings, which can be incorporated into $\mathcal{S}_c[\psi^{\dagger},\psi]$.  So after integrating out the $\mathbf{L}$ field we find:
\begin{eqnarray}
	Z &=& \int \mathcal{D}\mathbf{n}\mathcal{D}\alpha\mathcal{D}\psi^{\dagger}\mathcal{D}\psi\;\delta(\mathbf{n}^2-1) 
e^{-\mathcal{S}_{KLM}[\mathbf{n},\alpha, \mathbf{s}_c]-\mathcal{S}_c[\psi^{\dagger},\psi]-\mathcal{S}_{B}^{\prime}[\mathbf{n}]} \nonumber\\
	\mathcal{S}_{KLM}[\mathbf{n}, \alpha,\mathbf{s}_c] 
		&=& const_2 + \frac{c}{2g}\int_0^{\beta}d\tau\int d^dx \Bigg[ \left(\nabla\mathbf{n}(\vec{x},\tau)\right)^2 
+ \frac{1}{c^2}\left(\frac{\partial\mathbf{n}(\vec{x},\tau) }{ \partial\tau}\right)^2 \Bigg]  \nonumber\\
	&\quad& +\lambda\int d^d xd\tau\left( \mathbf{s}_c(\vec{x},\tau)\cdot\mathbf{n}(\vec{x},\tau)
\times\frac{\partial\mathbf{n}(\vec{x},\tau)}{\partial\tau}\right) \nonumber\\
	&&+ \frac{1}{2gc} \int_0^{\beta}d\tau\int d^dx \; \alpha^2(\vec{x},\tau) 
\end{eqnarray}
The constants $g$, $c$, and $const_2$ are defined exactly as before, and the new Kondo coupling constant is:
\begin{equation}
	\lambda \equiv \frac{iJ_K}{4dI_1 a^d }
\end{equation}
Finally, we integrate out the $\alpha$ field which only contributes the same constant as before.  The final result is:
\begin{eqnarray}
	Z &=& \int \mathcal{D}\mathbf{n}\mathcal{D}\psi^{\dagger}\mathcal{D}\psi\;\delta(\mathbf{n}^2-1) 
e^{-\mathcal{S}_{QNL\sigma M}-\mathcal{S}_{K}[\mathbf{n}, \mathbf{s}_c]-\mathcal{S}_c[\psi^{\dagger},\psi]
-\mathcal{S}_{B}^{\prime}[\mathbf{n}]} \nonumber\\
\\
	\mathcal{S}_{\text{QNL$\sigma$ M}}[\mathbf{n}]
		&=& const_3 + \frac{c}{2g}\int_0^{\beta}d\tau\int d^dx \Bigg[ \left(\nabla\mathbf{n}(\vec{x},\tau)\right)^2 \nonumber\\
		&&+ \frac{1}{c^2}\left(\frac{\partial\mathbf{n}(\vec{x},\tau) }{ \partial\tau}\right)^2 \Bigg] 
\label{QNLsM}\\
	\mathcal{S}_{K}[\mathbf{n},\mathbf{s}_c] &=& \lambda\int d^d xd\tau\left( \mathbf{s}_c(\vec{x},\tau)
\cdot\mathbf{n}(\vec{x},\tau)\times\frac{\partial\mathbf{n}(\vec{x},\tau)}{\partial\tau}\right)
\label{Kondo-QNLsM}\\
	\mathcal{S}_c[\psi^{\dagger},\psi] &=& \int d^dK d\varepsilon\sum_{\sigma}\psi^{\dagger}_{\sigma}
(\vec{K},i\varepsilon)(i\varepsilon-\xi_K)\psi_{\sigma}(\vec{K},i\varepsilon) + u\int \psi^4
\end{eqnarray}
Note that terms from $\mathbf{s}_c^2$ have been absorbed in $u$ and $\xi_K$.  
This completes the mapping from the microscopic Kondo Lattice Hamiltonian
to the effective field theory, as claimed earlier.

Now that we have demonstrated this mapping in detail, we are confronted with
executing the 
RG analysis.
The next section is devoted to this theoretical development.

\section{Renormalization group analysis and the antiferromagnetic phase with Kondo breakdown}

In the previous section we presented the details on the 
construction of the effective field
theory corresponding to the antiferromagnetic 
Kondo (Heisenberg) lattice.  The end result is a
QNL$\sigma$M coupled to itinerant electrons with a Fermi surface.  
An RG analysis of this field theory is much more involved than 
field theories typically encountered in high energy physics.
These complications stem from the existence 
of a Fermi surface~\cite{Shankar1994}.
We wish to treat the gapless fermions and bosons on an equal footing
instead of integrating out the fermions first;
doing the latter would introduce undesirable
non-analyticities, as first pointed out by \cite{Vojta1997, Abanov2004}.

\begin{figure}[hbtp]
   \centering
   \includegraphics[width=3in]{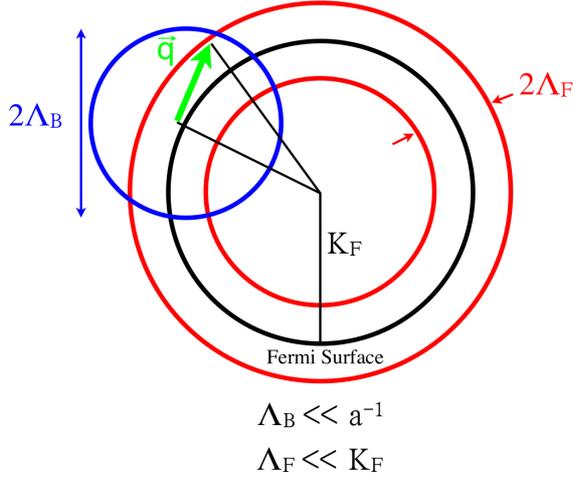}
   \caption[Renormalization group for mixed fermions and bosons.]
   {
Kinematics of the RG rescaling \cite{Yamamoto10}
for a system containing both gapless 
bosons, whose momenta are centered around a point in the momentum space,
and gapless fermions, whose momenta are confined to the vicinity
of a Fermi surface. $\Lambda_B$ and $\Lambda_F$ are the cut-off momenta
for the bosons and fermions, respectively. $a$ is the lattice constant
in real space.
}
\label{fig:boson.fermion.scaling}
\end{figure}

We have developed an RG approach for mixed fermion-boson
theories \cite{Yamamoto10}, building on the method for
purely-fermionic problems ~\cite{Shankar1994}. 
The approach is not technically difficult, but the subtleties of the
scaling procedure are not straightforward either. 
Our method meshes the scaling of bosons, which takes place along 
all directions in the 
momentum space with respect to a point in momentum space,
and that of fermions, which involves only one direction locally 
perpendicular to the Fermi surface. This is illustrated in 
Fig.~\ref{fig:boson.fermion.scaling}. For a long exposition 
of this scaling procedure, we refer the 
%\sout{readers} 
reader 
to our recent paper
\cite{Yamamoto10}. Here, we wish to emphasize only one particularly
pertinent aspect. Our mixed fermion-boson approach is 
%\sout{particularly} 
simplified in 
the antiferromagnetic case considered here
 by the fact that the bosonic modes of the QNL$\sigma$M
action [Eq.~(\ref{QNLsM})] has a dynamic exponent $z=1$.

%%%%%%%%%%%

\subsection{The case of Fermi surface not intersecting the antiferromagnetic zone boundary}

The analysis based on the combined fermion-boson RG 
%\sout{was} 
has been
discussed 
elsewhere \cite{Yamamoto07}. 
The effective Kondo coupling, expressed in the QNL$\sigma$M representation 
by Eq.~(\ref{Kondo-QNLsM}), is marginal at both the tree, one-loop, and infinite-loop
levels. The tree level and one-loop analyses directly use the procedure of 
Ref.~\cite{Yamamoto10}, while the infinite-loop analysis was based on a decomposition
of the Fermi surface into patches.

Here we present an alternative infinite-loop analysis without appealing to patching arguments.
The goal of this analysis
is to establish a type of Migdal's Theorem which states that the tree-level
result is the entire story.  Since we found marginality at the tree-level, this
is the exact answer to all orders in the limit where $\Lambda/K_F \to 0$.

For concreteness, we consider a spherical Fermi surface although generalizations to non-nested generic Fermi
surfaces can be readily made. 
For a spherical Fermi surface we can write the Fermi momentum integral in terms of spherical coordinates:
\begin{eqnarray}
	\int d^dK 
	&=& \int_{K_F - \Lambda}^{K_F + \Lambda} K^{d-1}dK \int d^{d-1}\Omega_K \nonumber \\
	&=& \int_{- \Lambda}^{\Lambda} (k+K_F)^{d-1} dk \int d^{d-1}\Omega_K
\end{eqnarray}
The most relevant part of the above is
\begin{eqnarray}
	\int d^dK 
	&=& K_F^{d-1} \int_{- \Lambda}^{\Lambda}  dk \int d^{d-1}\Omega_K
\end{eqnarray}
Now the kinetic part of the fermions can be written,
\begin{eqnarray}
	\mathcal{S}_c
	&=& K_F^{d-1} \int dk d^{d-1}\Omega_K d\varepsilon \; \psi^{\dagger} (i\varepsilon - v_F k) \psi
\end{eqnarray}
We define new dimensionless variables:
\begin{eqnarray}
	\varepsilon &=& \Lambda\bar{\varepsilon} \\
	k &=& \Lambda\bar{k} \\
	\Omega_K &=& \bar{\Omega}_{K} \\
	K_F^{d-1}\Lambda^{3} \psi^{\dagger}\psi &=& \bar{\psi}^{\dagger}\bar{\psi}
\end{eqnarray}
Note that the angular components of fermionic momenta are untouched.  We now have:
\begin{eqnarray}
	\mathcal{S}_c
	&=&  \int d\bar{k} d^{d-1}\bar{\Omega}_{K} d\bar{\varepsilon} \; \bar{\psi}^{\dagger} (i\bar{\varepsilon} - v_F \bar{k}) \bar{\psi}
\end{eqnarray}
The important difference from the aforementioned patching argument is that now the fermionic fields contain factors of $K_F$.  Plugging this into the Kondo coupling we find (note that the $\text{QNL$\sigma$M}$ rescaling is identical to what was done in Ref.~\cite{Yamamoto07})
\begin{eqnarray}
	\mathcal{S}_{K} 
	&=& K_F^{1-d}\Lambda^{1+1+d+1-3+1-(d+3)/2}\lambda_{\perp} \int d\bar{k}d^{d-1}\bar{\Omega}_{K} d\bar{\varepsilon} d^d\bar{q} d\bar{\omega} [\bar{\psi}^{\dagger}\bar{\psi}(\bar{\omega} \bar{\pi})] \\
	&=& \frac{\Lambda^{(d-1)/2}}{K_F^{d-1}} \lambda_{\perp} \int d\bar{k}d^{d-1}\bar{\Omega}_{K} d\bar{\varepsilon} d^d\bar{q} d\bar{\omega} [\bar{\psi}^{\dagger}\bar{\psi}(\bar{\omega} \bar{\pi})] 
\end{eqnarray}
For $d=2$ we have
\begin{eqnarray}
	\frac{ \mathcal{S}_{K} }{ \mathcal{S}_c } 
	&\propto& \frac{\sqrt{\Lambda} }{K_F} = \frac{1}{\sqrt{N_{\Lambda}}}\frac{1}{\sqrt{K_F}}
\end{eqnarray}
In the non-spin-flip channel:
\begin{eqnarray}
	\Gamma_{z} 
		&=& K_F^{1-d} \Lambda^{1+1+d+1+d+1-3+1-2((d+3)/2)} \lambda_{z} \int d\bar{k}d^{d-1}\bar{\Omega}_{K} d\bar{\varepsilon} d^d\bar{q}_1 d\bar{\omega}_1 d^d\bar{q}_2 d\bar{\omega}_2 [\bar{\psi}^{\dagger}\bar{\psi}(\bar{\omega} \bar{\pi}\bar{\pi})]  \nonumber \\
		&=& \frac{\Lambda^{d-1} }{K_F^{d-1} } \lambda_{z} \int d\bar{k}d^{d-1}\bar{\Omega}_{K} d\bar{\varepsilon} d^d\bar{q}_1 d\bar{\omega}_1 d^d\bar{q}_2 d\bar{\omega}_2 [\bar{\psi}^{\dagger}\bar{\psi}(\bar{\omega} \bar{\pi}\bar{\pi})]  
\end{eqnarray}
For $d=2$,
\begin{eqnarray}
	\frac{ \Gamma_{z} }{ \mathcal{S}_c } &\propto& \frac{\Lambda}{K_F} = \frac{1}{N_{\Lambda}}
\end{eqnarray}

We have thus shown that for any $d>1$ the Kondo vertex will have associated with it positive 
powers of $1/N_{\Lambda}$.
Because of this, as the number of powers of $J_K$ increases, so does the suppression
factor $1/N_{\Lambda}$. The only exception is for a series of diagrams corresponding to 
a chain of particle-hole bubbles in the spin-flip channel \cite{Yamamoto07}.
Because the poles are located on the same side of the real axis, they make no contribution
to the beta function \cite{Yamamoto07,Shankar1994}.
Thus, the tree-level result is the whole story, and the Kondo coupling is exactly marginal.

The RG result is further corroborated by a large-N calculation for the 
conduction electron Green's function, which 
%explicitly shows 
does not contain a pole and, correspondingly, the Fermi surface is 
small \cite{Yamamoto07}. 

%%%%%%%%%%%%
\subsection{The case of Fermi surface intersecting the antiferromagnetic Brillouin zone boundary}
\label{AFBZIntersection}
%%%%%%%%%%%%

We now turn to the case where the Fermi surface intersects the 
%antiferromagnetic Brillouin zone (AFBZ)
AFBZ boundary.
In this case, the linear coupling 
$\vec{n}\cdot\vec{s}_c $ between the local moments and conduction
electron spin cannot be neglected. 
Until now, we have only considered the term 
$\vec{L}\cdot\vec{s}_c$
because our assumption has been that the Fermi
surface does not intersect the AFBZ boundary, {\it i.e.} $Q > 2K_F$.
See Fig~\ref{fig1}a and the comments following
equation (\ref{eq:AFQNLSMCoupling}).
When $Q < 2K_F$, the conduction electrons see the AF order parameter
of the local moments as a staggered scattering potential, 
resulting in a reconstruction of their Fermi
surface.  The hot spots of the Fermi surface therefore become gapped out,
as shown in Fig.~\ref{fig1}b. 
The effective coupling \cite{Yamamoto08} 
between the reconstructed quasiparticles and 
the $\vec{n}$ field 
involves a coherence factor containing 
an additional factor of $q$ (measured w.r.t. $\vec{Q}$).
This linear-momentum suppression factor survives
beyond the mean-field treatment of the conduction electron band,
as dictated by Adler's Theorem.
Indeed, it can be viewed as a kinematic suppression 
similar to the deformation potential problem 
of the electron-phonon system~\cite{Schrieffer1995}.

From the RG perspective, the additional factor of $q$ represents
a decrease in
the dimension of the coupling which has the same effect as 
a derivative coupling:
$\vec{s}_c \cdot \vec{n}_q \to q a^{\dagger} \vec{\sigma} a 
\cdot \vec{n}_q$.
Now the conduction electron spin is coupled directly to $\vec{n}_q$ which has 
dimension $\left[ \pi_q \right] = -d-1$, but the additional factor of $q$ 
brings
the dimension to $\left[ q\; \vec{n}_q \right] = -d$ which has the same
value as the vector field $\left[ \vec{\varphi}_q \right] = -d$ 
we considered earlier
for the case where the Fermi surface does not intersect the AFBZ boundary
\cite{Yamamoto07}.
Therefore, our previous result on the marginality of the Kondo coupling 
is not spoiled 
when the Fermi surface intersects the magnetic zone boundary.

The marginal nature of the Kondo coupling implies that the effective 
Kondo coupling does not flow towards strong coupling. Correspondingly,
there is no Kondo singlet formation in the ground state, and no Kondo
resonances in the single-electron excitation spectrum. The absence of 
the Kondo resonance implies that the local moments remain charge-neutral,
and the Fermi surface 
%will be small, 
is determined by the conduction electrons alone; such a Fermi surface
is called small, and the antiferromagnetic phase is named ${\rm AF_S}$.
This is to be contrasted with what happens in an antiferromagnetic
phase in the presence of Kondo resonances, in which the 
Fermi surfaces are specified by the hybridizing conduction electrons 
and the itinerant $f$-electrons that describe the Kondo resonances;
such a Fermi surface is called large, and this antiferromagnetic phase
is named ${\rm AF_L}$. Note that the presence of antiferromagnetic 
order can be incorporated in these definitions through an appropriate
symmetry-breaking field ~\cite{Pivovarov.04}.

\section{Global phase diagram of the antiferromagnetic Kondo lattice}

\begin{figure}[hbtp]
   \centering
   \includegraphics[width=3in]{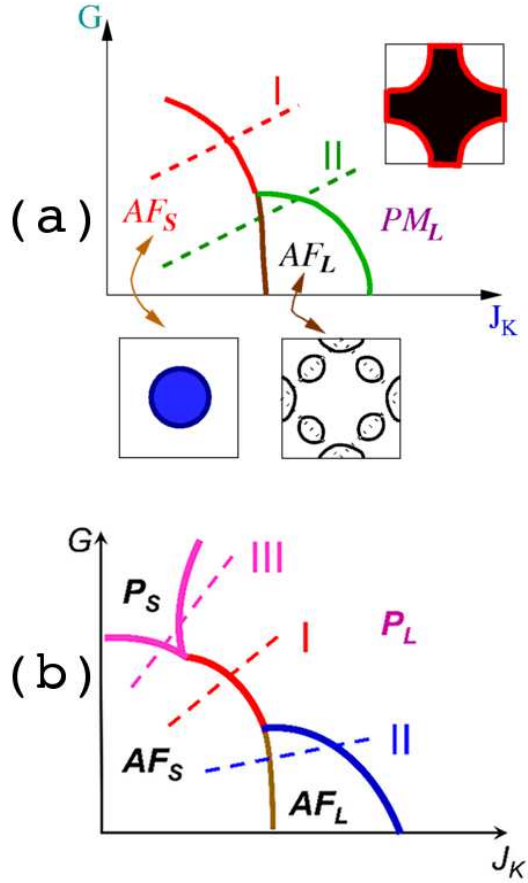}
   \caption[Global phase diagram of antiferromagnetic heavy fermions.]
   {
(a) Global phase diagram with the illustrative Fermi surfaces. 
(From \cite{Si06}.)
(b) Global phase diagram showing the three different types of trajectories
for quantum phase transitions. This phase diagram has a remarkable 
correspondence with what happens in heavy fermion metals.
(From \cite{Si10}.)
}
\label{fig:globa.pd}
\end{figure}

The establishment of the ${\rm AF_S}$ phase, along with the heavy-fermion
${\rm P_L}$ phase~\cite{Auerbach-prl86, Millis87}, provide two anchoring 
points in
the heavy-fermion
phase diagram at $T=0$. In order to connect these two phases, 
Ref.~\cite{Si06} introduced a two-parameter phase diagram, 
$(J_K, G)$, 
which is shown in Fig.~\ref{fig:globa.pd}a. 
In units of conduction electron bandwidth, $J_K$ is the Kondo coupling
while $G$ corresponds to magnetic frustration or reduced dimensionality.
The latter is a measure of the quantum fluctuations amongst the local-moment
degrees of freedom.
It was also discussed in Ref.~\cite{Si06}
that, for sufficiently large $G$, the conventional 
AF state of the local-moment component will yield to paramagnetic 
states which either preserve spin-rotational invariance (gapped 
or gapless spin liquid) or break it (spin Peierls). 
Inspired by the recent developments
in doped ${\rm YbRh_2Si_2}$, this part of the phase diagram is explicitly
included in the global phase diagram \cite{Si10} as shown in 
Fig.~\ref{fig:globa.pd}b. Related considerations on the global phase diagram
are also being made in Ref.~\cite{Coleman10}.

%%%%%%%%%%%

This global phase diagram can be described in terms of three possible
trajectories that connect the ${\rm AF_S}$ phase and ${\rm P_L}$ phase,
as specified in Fig.~\ref{fig:globa.pd}b. 
%This global phase diagram contains three routes for a system to
%go from the ${\rm AF_S}$ phase to the ${\rm P_L}$ phase.
Trajectory $I$ goes directly between the ${\rm AF_S}$ and ${\rm P_L}$ 
phases, giving rise to a local quantum critical point:
the collapse of Kondo resonances occurs at the magnetic quantum
critical point -- in the notation of 
Refs.~\cite{Si-Nature, Si-prb03,SiSmithIngersent99},
$\delta_c$ and $\delta_{loc}^c$ are located at the same place.
As the Kondo effect is continuously broken down at the AF
QCP, there is a sudden jump between the large Fermi surface 
and the small one ~\cite{Si-Nature,Coleman01,Si-prb03,SiSmithIngersent99},
and the Kondo-breakdown scale $E_{\mathrm{loc}}^*$ vanishes at the 
QCP. Correspondingly,
the residues associated with both the small
and large Fermi surfaces vanish as the QCP is approached
from either
side~\cite{Si-Nature, Coleman01, Si-prb03, Si06, Senthil06}.
Our considerations of the global phase diagram puts local
quantum criticality in a larger perspective.

Along Trajectory $II$, the transition between the ${\rm AF_S}$ and 
${\rm P_L}$ phases involve an intermediate ${\rm AF_L}$ phase. This is
the SDW state of the heavy quasiparticles of the 
${\rm P_L}$ phase. The magnetic-to-paramagnetic transition 
is of the SDW type~\cite{Hertz1976,Millis93,Moriya1985}.
A Kondo breakdown transition can still take place at the
${\rm AF_L}-{\rm AF_S}$ boundary~\cite{Si06,Yamamoto07};
in typical cases, this corresponds to a 
a Lifshitz transition with a change of Fermi surface topology.

Finally, along Trajectory $III$, the transition goes through 
the intermediate ${\rm P_S}$ phase. This is a paramagnetic phase
with a small Fermi surface. The small-to-large Fermi surface 
transition could be either a 
spin-liquid~\cite{Senthil04, Anderson10}
to heavy-Fermi-liquid
QCP,
or a 
spin-Peierls 
to heavy-Fermi-liquid 
QCP~\cite{Pivovarov.04}.

There is overwhelming evidence that pure ${\rm YbRh_2Si_2}$ at ambient 
pressure displays a type-I quantum phase transition. Recently,
Friedemann {\it et al.} \cite{Friedemann09} showed that 
%with 
enough Co-doping 
%relatively small amount of 
(of nominally 3\% or more),
% Co- doping, 
which introduces
positive chemical pressure,
turns the transition into one whose properties are largely compatible with
the type-II transition. Preliminary studies have shown that 
a similar effect arises in pure ${\rm YbRh_2Si_2}$ under 
a sufficiently large pressure
\cite{Tokiwa.09}.
A relatively small amount of (nominally 2.5\%) 
Ir-doping, which introduces negative chemical pressure, 
retains the type-I transition \cite{Friedemann09}.
By contrast, a larger negative chemical pressure, corresponding to a 
larger amount (nominally 6\% or more) Ir- doping, turns the transition
into one that is compatible with a type-III transition. A similar behavior
has also been seen in Ge-doped ${\rm YbRh_2Si_2}$ \cite{Custers10}.
Experimentally, there is an indication that the regime corresponding to 
$P_S$ has non-Fermi liquid behavior, raising the exciting possibility
that this is in fact a non-Fermi liquid phase.

\section{Kondo insulators}

Heavy fermion metals involve localized magnetic moments and a partially-filled
conduction-electron band. This can be generically modeled in terms of a
Kondo lattice Hamiltonian comprising two specifies of electrons:
a lattice of spin-$1/2$ local moment, with one per unit cell; and
a band of conduction electrons with a filling of
$0<x<1$ electrons per unit cell.

When $x=1$, we have an even number of electrons per unit cell and an insulator
becomes a natural possibility. In fact, the analogue of the ${\rm P_L}$ phase
of the $0<x<1$ case is the Kondo insulator\cite{Aeppli_Fisk92}.
For $x=1$, the Kondo-singlet formation
in the ground state induces delocalized $f$-electron quasiparticles,
which hybridizes with the conduction electron band and induces
a hybridization gap at the Fermi energy.
At zero-field, it is traditionally 
%considered 
believed
that the Kondo insulator
is the only possible phase. (By contrast, a large magnetic field can suppress
the Kondo effect altogether, inducing a magnetically ordered
metal \cite{Aeppli_Fisk92}.)

Here, we make the observation that the ${\rm AF_S}$ phase 
extensively described above
%earlier 
remains a stable phase for the $1+1$ ``Kondo-insulator filling.''
The exchange interaction 
%among 
between the local moments in this regime is
generically expected to be antiferromagnetic. In addition,
in the parameter regime specified by Eq.~(\ref{Kondo-AF-regime}),
our asymptotically exact RG analysis of the
Kondo lattice Hamiltonian in its QNL$\sigma$M representation continues
to apply, regardless of whether the underlying conduction electron
Fermi surface intersects the AFBZ boundary.
% of the AF Brizllouin zone.
To the same degree of confidence as in the case of generic 
%$1+1$ 
filling,
%case, 
the small-Fermi-surface paramagnetic ($P_S$) phase is also
expected to occur in the case of $1+1$ filling 
when the the parameter $G$ (the quantum fluctuations
of the local-moment component) becomes sufficiently large and 
for small
Kondo coupling.

The above assumes that there is no perfect nesting of the conduction electron
Fermi surface. A perfect nesting with 
$x=1$ would only occur for a square lattice with strictly no hopping beyond
nearest neighbors, and this is unlikely to be the case for rare-earth intermetallics.

\begin{figure}[hbtp]
   \centering
   \includegraphics[width=5in]{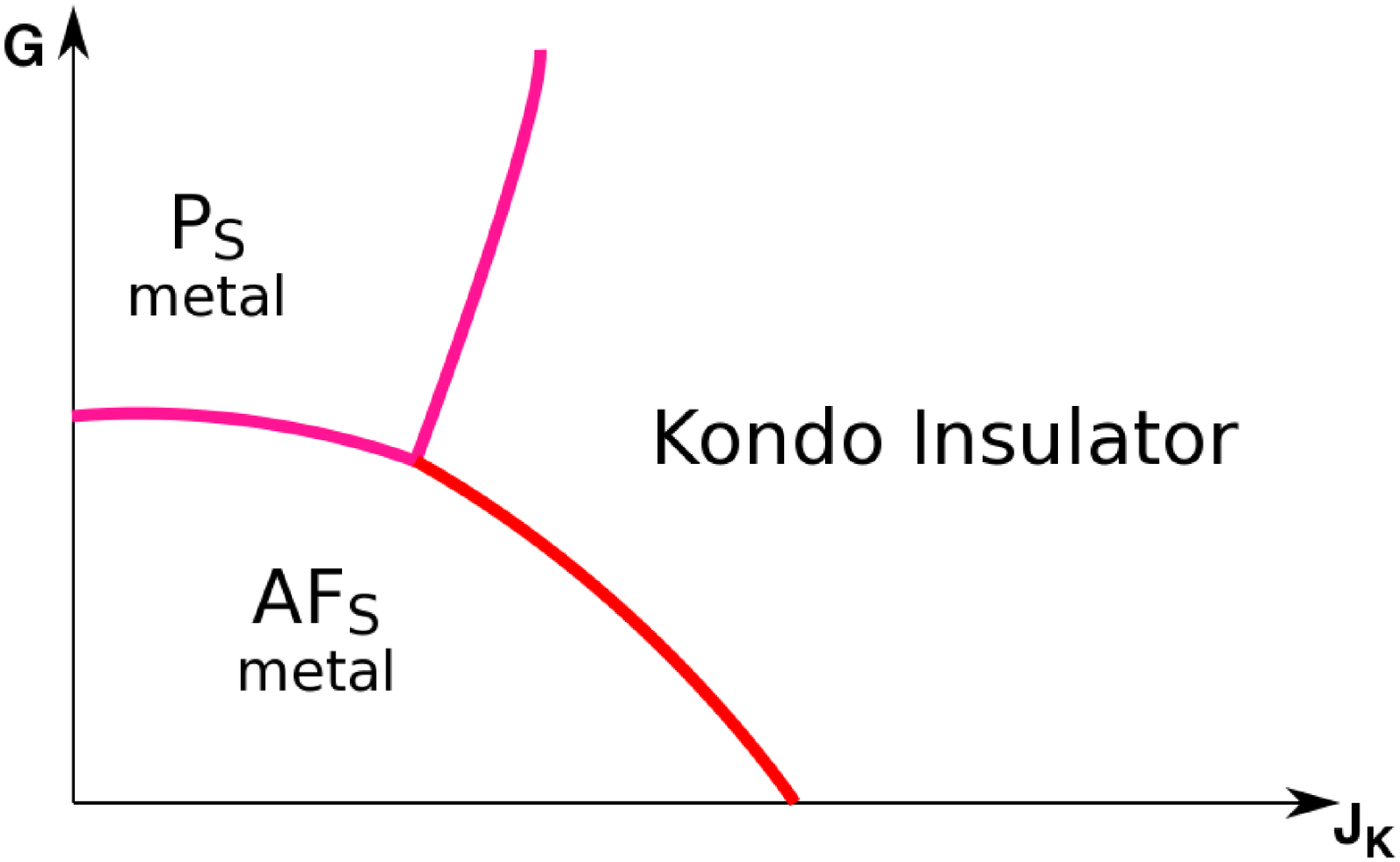}
   \caption[Global phase diagram for Kondo-insulator systems.]
   {
Global phase diagram for Kondo-insulator systems. The Kondo-insulator 
phase corresponds to a Kondo-singlet ground state, and is paramagnetic.
${\rm AF_S}$ describes an antiferromagnetic metal state
with a small Fermi surface and ${\rm P_S}$ a paramagnetic metal phase 
also with a small Fermi surface; each 
has the same
meaning as 
its counterpart
in Fig.~\ref{fig:globa.pd}.
}
\label{fig:ki}
\end{figure}

These considerations lead to a global phase diagram for Kondo insulators,
shown in Fig.~\ref{fig:ki}.
The list of materials believed to be Kondo insulators is by now relatively
large. It will be instructive to search for quantum phase
transitions out of a Kondo insulator and into either
the ${\rm AF_S}$ or ${\rm P_S}$ metallic phases.
While it is in general hard to predict precisely which trajectory a particular
tuning parameter -- such as pressure or chemical doping -- would correspond
to in such a global phase diagram, what is minimally required is to increase
the ratio of the AF RKKY interaction to the Kondo coupling.
For Ce-based systems, this happens with the application of
a negative (chemical) pressure. For Yb-based systems, on the other hand,
this is achieved by applying a positive (chemical) pressure.

%%%%%%%%%%%

\section{Connection to the holographic theory of non-Fermi liquid}

There has been considerable recent interest in applying the techniques
developed in string theory to many-body systems. These developments rely
on a duality between a $d$-dimensional strongly-coupled field theory 
and a quantum-gravity theory in a weakly curved $(d+1)$ dimensional
anti-de Sitter (AdS$_{d+1}$) spacetime. Non-Fermi liquid behavior
has been studied through a charged black hole in the AdS 
spacetime \cite{Lee09,Liu09,Cubrovic09,Faulkner09,Hartnoll10}.
In this approach, the fermionic self energy shows 
a marginal-Fermi-liquid-like form \cite{Varma89}. This has been
clarified as originating from an infrared fixed point characterized
by a $(0+1)$-dimensional conformal field theory \cite{Faulkner09}, 
which is dual to a near horizon geometry 
AdS$_2\times \ensuremath{\mathbb{{R}}}^{d-1}$.
The $(0+1)$-dimensional conformal field theory corresponds to a 
quantum impurity model. 

The factorization of temporal and spatial correlations become more
transparent in a ``semi-holographic'' 
formulation \cite{Faulkner:2010tq,Faulkner09}, 
in which a conduction-electron band
is introduced to be hybridized with the fermions in 
a strongly coupled field theory dual to the gravity theory. 
In turn, this allows an Anderson-lattice-model 
interpretation (see also Ref.~\cite{Sachdev:2010}),
in which the fermions of the
strongly coupled field theory is the analogue of the $f$-electron degree
of freedom of the Anderson lattice model. The 
relevance or irrelevance of the hybridization may then be 
linked to the Kondo screening or Kondo breakdown 
of a class of quantum impurity models
-- the Bose-Fermi Kondo model -- coupled
to a fermionic bath \cite{SmithSi99,Sengupta00,ZhuSi02,Zarand02}.
In the Kondo language, a breakdown of the Kondo effect,
occurring when the hybridization-induced Kondo coupling is 
irrelevant or marginal in the RG sense, leads to 
a power-law form of the conduction-electron self-energy whose 
exponent is positive; such a form of the self-energy 
appears in the non-Fermi liquid state of the holographic models 
\cite{Lee09,Liu09,Cubrovic09,Faulkner09,Hartnoll10}.
A Kondo-screened state, occurring when the hybridization-induced Kondo
coupling is relevant in the RG sense, gives rise to a pole 
in the conduction-electron self-energy. No such a form of the 
self-energy has been identified so far in the holographic models.

These considerations point to a linkage between the Kondo-breakdown
physics of the Kondo/Anderson lattice systems and the fate of the
non-Fermi liquid state dual to the gravity theory with a near-horizon
AdS$_2\times \ensuremath{\mathbb{{R}}}^{d-1}$ geometry. There could
be at least two possibilities for the latter.
One possibility is that the non-Fermi liquid state is unstable 
towards an antiferromagnetically-ordered state 
with a small Fermi surface discussed earlier,
but retains a similar form in the quantum-critical regime; 
this is the analogue of the Kondo-breakdown physics in the 
local quantum criticality formulation \cite{Si-Nature,Si-prb03}. 
Note that the entropy vanishes at the local quantum critical 
point \cite{Dai07}.
An alternative is that this state is unstable towards 
a paramagnetic phase with a small Fermi surface \cite{Sachdev:2010},
as occurring in one formulation of the Kondo 
breakdown~\cite{Senthil04,PaulPepinNorman07}.
In either case, these considerations suggest that holographic models
may contain an analogue of the paramagnetic phase with a large Fermi
surface, whose conduction-electron self-energy singularly depends on
energy (containing a pole) and smoothly depends on momentum.

With these considerations in mind, it will be instructive to study itinerant 
magnetic systems from the gravity side. A recent work has gone along 
this direction \cite{Iqbal10}.

\section{Conclusions}

Recent theoretical and experimental developments have opened up 
the issue of the global phase diagram in heavy fermion metals. 
We discussed some of the 
earlier developments that have led to the continued theoretical 
efforts on the global phase diagram, and the striking recent experiments that
point to the richness of the quantum
phase 
transitions between antiferromagnetic
and paramagnetic heavy fermion metals. 

We provided
the details of the asymptotically exact analysis of the Kondo lattice 
in the antiferromagnetic part of the phase diagram. General considerations
on the transitions from the Kondo-breakdown antiferromagnetic metal phase
with a small Fermi surface to the Kondo-screened paramagnetic heavy fermion
phase have motivated a 
global phase diagram, which is consistent with earlier 
studies of the Kondo breakdown effect from the paramagnetic side including the 
distinction between 
the local quantum critical point and the SDW
quantum critical point. Our considerations also put earlier formulations 
of the Kondo breakdown effect in 
the paramagnetic region into a general perspective.
Model studies that can cover all these different phases and transitions in 
one unified framework are called for.

We have also discussed the Kondo insulators along a similar line. We have proposed 
a related global phase diagram for such systems, which we hope will stimulate future
experiments.

Finally, we have discussed the connection between the Kondo-breakdown
local quantum criticality of Kondo lattice systems and 
the holographic non-Fermi liquid behavior.
Considerations of magnetic states and 
quantum magnetic transitions from a gravity dual are just beginning. Conversely,
studies of quantum critical behavior in magnetic many-body systems,
including heavy-fermion metals and insulators, may shed some light on gravity problems.

%%%%%%%%%%%

\begin{acknowledgements}
We would like to thank E. Abrahams, P. Coleman, P. Goswami, S. Friedemann,
S. Kirchner, K. Ingersent, N. Iqbal, H. Liu, H. v. L\"ohneysen, S. Paschen, 
F. Steglich, S. Wirth, L. Zhu, and J.-X. Zhu for useful discussions and/or
collaborations.
This work has been 
supported by
the NSF Grant No. DMR-1006985 and 
the Robert A. Welch Foundation Grant No. C-1411.
\end{acknowledgements}

%\pagebreak
%
%\bibliographystyle{spmpsci}
%\bibliography{paper}

\end{document}